\documentclass[letterpaper,iop]{emulateapj}
\pdfoutput=1 
\usepackage{thumbpdf} 
\usepackage{amsmath,amssymb} 
\usepackage{graphicx,mathptmx} 
\usepackage{natbib} 
\usepackage[usenames]{color} 
\usepackage{ifpdf}  
\usepackage[protrusion=true,expansion=true]{microtype} 
\usepackage{xcolor}[2007/01/21] 
\usepackage{refcount} 
\usepackage{textcase} 
\usepackage{epstopdf} 

\ifpdf
\pdfoutput=1 
\else

\fi

\newcommand{\beq}{
\begin{equation}
}
\newcommand{\eeq}{
\end{equation}
}
\newcommand{\beqa}{
\begin{eqnarray}
}
\newcommand{\eeqa}{
\end{eqnarray}
}

\newcommand{\kgfigbeg}[1]{
\begin{figure}
\hypertarget{#1}{}%
}
\newcommand{\kgfigend}[2]{
\label{f:#1}
\end{figure}
\bookmarksetup{color=[rgb]{0.54,0,0}}
\bookmark[rellevel=1,keeplevel,dest=#1]{Fig \ref*{f:#1}: {#2}}
\bookmarksetup{color=black} 
}

\newcommand{\kgfigstarbeg}[1]{
\begin{figure*}
\hypertarget{#1}{}%
}
\newcommand{\kgfigstarend}[2]{
\label{f:#1}
\end{figure*}
\bookmarksetup{color=[rgb]{0.54,0,0}}
\bookmark[rellevel=1,keeplevel,dest=#1]{Fig \ref*{f:#1}: {#2}}
\bookmarksetup{color=black} 
}

\newcommand{\kgtabbeg}[1]{
\begin{deluxetable}{#1}
}
\newcommand{\kgtabend}[2]{
\label{t:#1}
\end{deluxetable}
\bookmarksetup{color=[rgb]{0,0,0.54}}
\bookmark[
rellevel=1,
keeplevel,
dest=table.\getrefnumber{t:#1}
]{Table \ref*{t:#1}: #2}
\bookmarksetup{color=[rgb]{0,0,0}}
}

\newcommand{\kgtabstarbeg}[1]{
\begin{deluxetable*}{#1}
}
\newcommand{\kgtabstarend}[2]{
\label{t:#1}
\end{deluxetable*}
\bookmarksetup{color=[rgb]{0,0,0.54}}
\bookmark[
rellevel=1,
keeplevel,
dest=table.\getrefnumber{t:#1}
]{Table \ref*{t:#1}: #2}
\bookmarksetup{color=[rgb]{0,0,0}}
}

\newcommand{\units}[1]  {\ensuremath{\mathrm{{#1}}}}
\newcommand{\msun}     {\ensuremath{{{M}}_{\scriptscriptstyle \odot}}}

\newcommand{\lsunv}     {\ensuremath{{L}}_{{\scriptscriptstyle \odot},V}}
\newcommand{\kms}      {\ensuremath{~\mathrm{km~s^{-1}}}}

\newcommand{\msigma}   {\ensuremath{M}{--}\ensuremath{\sigma}}
\newcommand{\ml}       {\ensuremath{M}{--}\ensuremath{L}}
\newcommand{\mbh}      {\ensuremath{M}}

\providecommand{\ion}[2]{#1$\;$\textsmaller{\@Roman{#2}}}

\DeclareMathOperator{\sech}{sech}

\def\spose#1{\hbox to 0pt{#1\hss}}
\newcommand{\lta}{\mathrel{\spose{\lower 3pt\hbox{$\mathchar"218$}}
      \raise 2.0pt\hbox{$\mathchar"13C$}}}
\newcommand{\gta}{\mathrel{\spose{\lower 3pt\hbox{$\mathchar"218$}}
      \raise 2.0pt\hbox{$\mathchar"13E$}}}
\def\simlt{\mathrel{\rlap{\lower 3pt\hbox{$\sim$}}\raise 2.0pt\hbox{$<$}}}
\def\simgt{\mathrel{\rlap{\lower 3pt\hbox{$\sim$}} \raise 2.0pt\hbox{$>$}}}

\definecolor{KayhanCiteColor}{rgb}{0,0.08,0.35}
\definecolor{KayhanURLColor}{rgb}{0,0.08,0.35}
\definecolor{KayhanLinkColor}{rgb}{0,0.08,0.35}
\definecolor{KayhanPageColor}{rgb}{0,0.08,0.35}
\definecolor{medred}{rgb}{0.75,0.0,0.0}

\shorttitle{$M_\mathrm{BH}$ and the One Ring in NGC 3706}
\shortauthors{G\"{u}ltekin et al.}

\submitted{Accepted by The Astrophysical Journal}

\makeatletter
\ifx\undefined\hyperref
\usepackage[pdftitle={The Black Hole in NGC 3706},pdfauthor={Kayhan Gultekin},pdfsubject={Astrophysics},pdfkeywords={black hole physics --- galaxies: general --- galaxies: nuclei --- galaxies: bulges --- methods: statistical},letterpaper,linktocpage,colorlinks=true,linkcolor={KayhanLinkColor},urlcolor={KayhanURLColor},citecolor={KayhanCiteColor},hyperfootnotes=false]{hyperref}
\else\relax\fi
\makeatother
\usepackage[figure,figure*]{hypcap} 
\usepackage{atveryend} 
\usepackage[
  open,
  openlevel=3,
  atend
]{bookmark}[2010/04/08]  
\begin{document}

\label{firstpage}
 
\title{The Black Hole Mass and the stellar ring in NGC 3706\footnotemark[1]}
\footnotetext[1]{Based on observations made with the \emph{Hubble
Space Telescope}, obtained at the Space Telescope Science Institute,
which is operated by the Association of Universities for Research in
Astronomy, Inc., under NASA contract NAS 5-26555.  These observations
are associated with GO proposal 8687.}

\author{Kayhan G\"{u}ltekin\altaffilmark{1}}
\author{Karl Gebhardt\altaffilmark{2}}
\author{John Kormendy\altaffilmark{2}}
\author{Tod R.\ Lauer\altaffilmark{3}}
\author{Ralf Bender\altaffilmark{4}}
\author{Scott Tremaine\altaffilmark{5}}
\author{Douglas O.\ Richstone\altaffilmark{1}}
\affil{\altaffilmark{1}Department of Astronomy, University of Michigan, 500 Church Street, Ann Arbor, MI 48109
\href{mailto:kayhan@umich.edu}{kayhan@umich.edu}.}
\affil{\altaffilmark{2}Department of Astronomy, University of Texas, 1 University Station C1400, Austin, TX, 78712, USA}
\affil{\altaffilmark{3}National Optical Astronomy Observatory, P.O. Box 26732, Tucson, AZ, 85726, USA}
\affil{\altaffilmark{4}Universit\"ats-Sternwarte M\"unchen, Ludwig-Maximilians-Universit\"at, Scheinerstr.\ 1, D-81679 M\"unchen, Germany}
\affil{\altaffilmark{5}Institute for Advanced Study, Einstein Dr., Princeton, NJ 08540, USA}

\begin{abstract}
\hypertarget{abstract}{} We determine the mass of the nuclear black
hole (\mbh) in NGC 3706, an early type galaxy with a central surface
brightness minimum arising from an apparent stellar ring, which is
misaligned with respect to the galaxy's major axis at larger radii.
We fit new \emph{HST}/STIS and archival data with axisymmetric orbit
models to determine \mbh, mass-to-light ratio ($\Upsilon_V$), and dark
matter halo profile.  The best-fit model parameters with 1$\sigma$
uncertainties are $\mbh = (6.0^{+0.7}_{-0.9}) \times 10^8\ \msun$ and
$\Upsilon_V = 6.0 \pm 0.2\ \msun\ \lsunv^{-1}$ at an assumed distance
of 46 Mpc.  The models are inconsistent with no black hole at a
significance of $\Delta\chi^2 = 15.4$ and require a dark matter halo
to adequately fit the kinematic data, but the fits are consistent with
a large range of plausible dark matter halo parameters.  The ring is
inconsistent with a population of co-rotating stars on circular
orbits, which would produce a narrow line-of-sight velocity
distribution (LOSVD).  Instead, the ring's LOSVD has a small value of
$|V|/\sigma$, the ratio of mean velocity to velocity dispersion.
Based on the observed low $|V|/\sigma$, our orbit modeling, and a
kinematic decomposition of the ring from the bulge, we conclude that
the stellar ring contains stars that orbit in both directions.  We
consider potential origins for this unique feature, including multiple
tidal disruptions of stellar clusters, a change in the gravitational
potential from triaxial to axisymmetric, resonant capture and
inclining of orbits by a binary black hole, and multiple mergers
leading to gas being funneled to the center of the galaxy.
\bookmark[ rellevel=1, keeplevel,
dest=abstract
]{Abstract}
\end{abstract}

\section{Introduction}
\label{intro}

NGC 3706 is one of eight early-type galaxies identified as having a
surface brightness profile with a local minimum at the galaxy's center
\citep{2002AJ....124.1975L, 2005AJ....129.2138L} .  Because of
projection effects, a central decrease in surface brightness
corresponds to a more prominent decrease in luminosity density.  The
highest surface brightness in NGC 3706 appears to be associated with
an axisymmetric, highly flattened, edge-on ring, which is misaligned
from the galaxy's major axis at larger radii by about 42$^\circ$.  The other galaxies
with central minima (NGC 4073, NGC 4382, NGC 4406, NGC 6876, A260-BCG,
A347-BCG, and A3574-BCG) are consistent with having similar rings, but
the interpretation is less clear, in part because the rings in these
galaxies are not edge-on.  Thus NGC 3706 offers a unique look at this
phenomenon.

The most likely interpretation of the central stellar system in NGC
3706 is one called ``stellar-torus-added'' by
\citet{2002AJ....124.1975L}, in which stars are brought in close to
the galaxy's center, presumably from a merger and tidal stripping of a
dense stellar system \citep{2000ApJ...531..232H} or else from a merger
with a system containing enough gas to form stars in situ
\citep{2002AJ....124.1975L}.  The stellar tori may be related to the
nuclear stellar disks seen in M31 \citep[NGC
224][]{1993AJ....106.1436L}, NGC 4486A \citep{2005AJ....129.2636K},
and NGC 4382 \citep{2005AJ....129.2138L, 2011ApJ...741...38G} as well
as similar structures seen in several other galaxies
\citep[e.g.,][]{1994AJ....108.1579V, 1995AJ....110.2622L, laueretal96,
1998MNRAS.293..343V, 1998MNRAS.300..469S, 1998MNRAS.298..267V}.

Eccentric disks are thought to require a massive dark object to
maintain the aligned semimajor axes \citep{tremaine95}.  There is,
however, no alignment problem in a circular ring, and thus the presence 
of such a structure does not necessarily imply the existence of a central black hole.
\citet{2011arXiv1107.2923L} have suggested that the Galactic center
may have suffered a recent satellite-galaxy merger analogous to those
in the stellar-torus-added scenario.

The presence of an edge-on stellar ring in principle makes measurement
of a black hole mass more tractable in the same way that flat galaxy
disks make measurements of galactic potentials tractable---there is a
unique circular speed at any radius, which is directly related to the
gradient of the potential.  The black-hole mass measurement would be
made easier still if it were a filled-in edge-on disk rather than a
ring, since the circular speed could be measured at smaller radii.
Knowing the mass of the black hole in any galaxy with an obviously
unique stellar feature is also useful for determining the degree to
which black holes correlate with their host galaxy properties.  For
example, if the black hole in NGC 3706 and others like it tended to
have much larger or much smaller black holes than similarly sized
galaxies without these features, then it would be a hint that the
tidal stripping or dissipative merging event that is suspected of
producing the ring is not connected to what drives the
galaxy-black-hole correlations.  What we find below is that the
black hole mass is small-to-average compared to the predictions of the
\msigma\ and \ml\ relations.

Section \ref{observations} has a brief description of NGC 3706 plus a
description of the observations and data new to this work as well as
previously published data that we use in our kinematic modeling.  In
Section \ref{modeling} we describe our kinematic modeling and results,
including our estimates of the black hole mass, mass-to-light ratio,
and dark matter halo properties.  We end with a brief discussion of
the implications of our results including speculations on the origin of a counter-rotating ring in Section \ref{discussion}.  We
adopted a distance of $46\ \units{Mpc}$ to NGC 3706 based on
its recession velocity with respect to the cosmic microwave background
and an assumed Hubble constant of $H_0 = 70\ \units{km\ s^{-1}\
Mpc^{-1}}$ \citep{laueretal07b}.  All quantities in this paper are
scaled to this distance.

\section{Observations of NGC 3706}
\label{observations}

NGC 3706 is in \citet{faberetal89} group 242.  It is classified as an
S0 based on images that show the appearance of diffuse starlight
outside of an obvious bulge component.  It may not be a true S0 galaxy
but only appear so because the extended diffuse starlight may be
``Malin shells'' resulting from recent merger activity
\citep{1980Natur.285..643M, 1984ApJ...279..596Q}.  It is bulge
dominated (type I in the \citealt{1995A&A...293...20S} classification
according to \citealt{1998A&AS..131..265S}).
\citet{1998A&AS..131..265S} find a bulge-to-total ratio of $B/T =
0.94$.  The disk component of their fits has a scale length of
2.82\ kpc---much smaller than their fitted bulge scale length of 8.89
kpc---and has $V$ magnitude $-19.25$.  Because the galaxy is bulge
dominated, the disk component may not arise from a true disk but
merely from deviations from the $r^{1/4}$ surface brightness profile,
potentially related to the stellar ring.  The disk component derived
by \citet{1998A&AS..131..265S} has an inclination of $67^\circ$ and is
typically $1.5 \mathrm{mag}\;\mathrm{arcsec}^2$ or more fainter than
the bulge; so its existence and properties are sensitive to small
deviations from the assumed bulge profile.  In any case, the
contribution of any such disk to the total luminosity is small and
will not affect our interpretation.  We adopt a bulge magnitude of
$M_V = -22.26$ \citep{1998A&AS..131..265S, laueretal05}.  Without the
stellar ring, NGC 3706 appears to have a core surface brightness
profile.  That is, the surface brightness profile has a form $I(r)
\sim r^{-\gamma}$ that transitions from a steep power law with $\gamma
\sim 1$ at large radii to a shallow core with $\gamma < 0.3$
\citep{1995AJ....110.2622L}. The conclusion that NGC 3706 has a core,
however, might depend on the model-dependent details of bulge-ring
decomposition \citep{laueretal05}.

NGC 3706's stellar ring has a total luminosity of about $1.3 \times
10^{8} \lsunv$. It is seen edge-on with a position angle (PA) of
$115^\circ$.  The ring has a maximum in surface brightness at
approximately 0\farcs13 from the center.  The ring has an inner edge
of about 0\farcs1 and is detectable to at least 0\farcs4 at which
point there is not enough contrast to separate it from the rest of the
stars in the galaxy.  Along its minor axis, the ring is unresolved at
\emph{Hubble Space Telescope} (\emph{HST}) resolution.
\citet{2002AJ....124.1975L} argued that the limb brightening, or
ansae, seen in the stellar feature at \emph{HST} resolution indicates
that it is a ring, i.e., there is a minimum in the stellar density at
zero radius, rather than a filled disk.

In order to model the kinematics of the galaxy and detect any central
black hole, we need high spatial resolution photometry and
spectroscopy of the center of the galaxy plus wide-field photometry
and spectroscopy.  Because of the high surface brightness and thinness of the ring, it makes for an excellent \emph{HST} target.  In
the following subsections, we describe the data we used in our
kinematic modeling.  This includes previously published Wide Field Planetary Camera 2 (WFPC2) data,
archival ground-based spectroscopic data, and previously unpublished
Space Telescope Imaging Spectrograph (STIS) data.

\subsection{Photometry}
\label{photometry}
Our \emph{HST} photometry comes from \citet{2002AJ....124.1975L} WFPC2
F555W imaging (GO-6597, PI: D.\ Richstone), which is the only data set
to reveal the stellar ring (Fig.\ \ref{f:slits}).  Reduction of the
NGC 3706 images is described in detail in \citet{2002AJ....124.1975L}.
In brief, the data were taken in a 500 s exposure with a square $2\
\times\ 2$ pattern of half-pixel steps to provide Nyquist sampling.
The images were combined using the Fourier method of
\citet{1999PASP..111..227L}.  The PSF was assembled from observed stars obtained close in time to the imaging observations \citep{2002AJ....124.1975L, 2005AJ....129.2138L}.   The PSF was deconvolved using 40
iterations of the Lucy--Richardson deconvolution
\citep{1974AJ.....79..745L, 1972JOSA...62...55R} prior to the
subsequent analysis.  The WFPC2 data are useful only out to about
100\arcsec; so we add ground-based data provided by
\citet{1994A&AS..104..179G} and \citet{1994MNRAS.270..523C} to extend
our radial coverage.

\kgfigstarbeg{slits}
\centering
\includegraphics[width=0.3\textwidth]{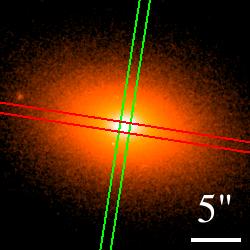}
\includegraphics[width=0.3\textwidth]{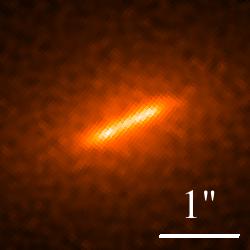}
\includegraphics[width=0.3\textwidth]{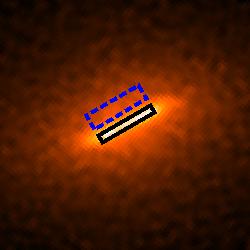}
\caption{\emph{HST}/WFPC2 F555W images of NGC 3706.  The left panel
  shows the galaxy on large scales with overlays showing the approximate
  slit positions due to \citet{1994MNRAS.270..523C} for the
  ground-based kinematic data that we use.  The center and right
  panels are from the deconvolution due to
  \citet{2002AJ....124.1975L}.  The center panel shows the presence of
  the stellar ring and some of the limb brightening.  The
  black and blue boxes in the right panel (at the same scale) show the
  approximate positions of the STIS slits for the central major axis
  and off-center major axis pointings, respectively.  All images are
  oriented as north-up, east-left.  The scale is shown by the labeled
  horizontal bar.}
\kgfigstarend{slits}{Galaxy images showing slit placement.}

\subsection{Spectroscopy}
\label{spectroscopy}

As part of GO-8687 (PI: J.\ Kormendy), we observed \ion{Ca}{2} triplet
absorption from NGC 3706 with STIS on \emph{HST}.  The spectrograph
was operated with the G750M grating for two different positions and slit
widths.  The STIS $52\arcsec\times0\farcs1$ slit was placed directly
on the WFPC2-measured location of the stellar ring to a requested
PA accuracy of $1^\circ$.  A second STIS pointing was
positioned with the $52\arcsec\times0\farcs2$ slit at a PA
parallel to the stellar ring but offset perpendicularly by 0\farcs25
to the North.  Figure \ref{f:slits} contains a diagram of the slit positions.

At the ring position, we obtained 6 exposures at 3 dither positions
for a total exposure of 15,174\ \units{s}.  At the offset position we
obtained 2 exposures at 2 dither positions for a total exposure of
5,648\ \units{s}.  The STIS CCD has a 1024$\times$1024 pixel format,
a readout noise of $\sim 1\ e^-\ \mathrm{pix}^{-1}$, and a gain of 1.0
without on-chip binning.  The wavelength range of the spectra was
8257--8847\ \AA, with a reciprocal dispersion of
0.554\ \AA\ pix$^{-1}$ and a spatial scale of 0\farcs05071
pix$^{-1}$ for G750M at 8561 \AA.

Our reduction of the STIS data followed the standard pipeline.  We
extracted raw spectra from the multi-dimensional file and then
subtracted a constant fit to the overscan region to determine the bias
level.  We used our iterative \emph{self-dark} technique
\citep{pinkneyetal03} to account for dark current.  This technique
takes into account the STIS CCD's warm and hot pixels that change on
timescales of roughly a day.  Then we flat-field, dark-subtract, and
shift the spectra vertically to a common dither, combine, and rotate.
We then extract one-dimensional spectra using a bi-weight combination
of rows.

We extracted line-of-sight velocity distributions (LOSVDs) from the
spectra, which were reduced as described in
\citet{2009ApJ...695.1577G}.  We deconvolved the observed galaxy
spectrum using a library template made of standard stellar spectra
\citep{gebhardtetal03, pinkneyetal03}.  Table \ref{t:n3706stisdata}
lists the Gauss-Hermite moments of the extracted velocity profiles;
Figures \ref{f:losvd0} and \ref{f:losvd1} plot the observed LOSVDs and
best-fit model in each bin; and Figure \ref{f:spec} contains a plot of
the mean velocity, $V$, and velocity dispersion, $\sigma$, as a
function of radius.  

The magnitude of the uncertainties in the LOSVD bins and of the
uncertainties in the Gauss-Hermite moments is neither uniform nor
symmetric.  The uncertainties are much larger for the $R =
-0\farcs075$, $0\farcs25$ spatial bins on the major axis and for the
$R=-0\farcs075$ on the offset slit position.  Although the surface
brightness is symmetric, the signal-to-noise of the extracted LOSVDs
is not symmetric because of changes in the equivalent width of the
absorption lines and kinematics.  The asymmetry in equivalent width
could be a result of an unusual stellar population gradient or
non-axisymmetric kinematics, though we find below that our
axisymmetric models produce an acceptable fit, and if the
gravitational potential is axisymmetric, any asymmetry should rapidly
be phase-mixed away.

\kgfigstarbeg{losvd0}
\includegraphics[width=\textwidth]{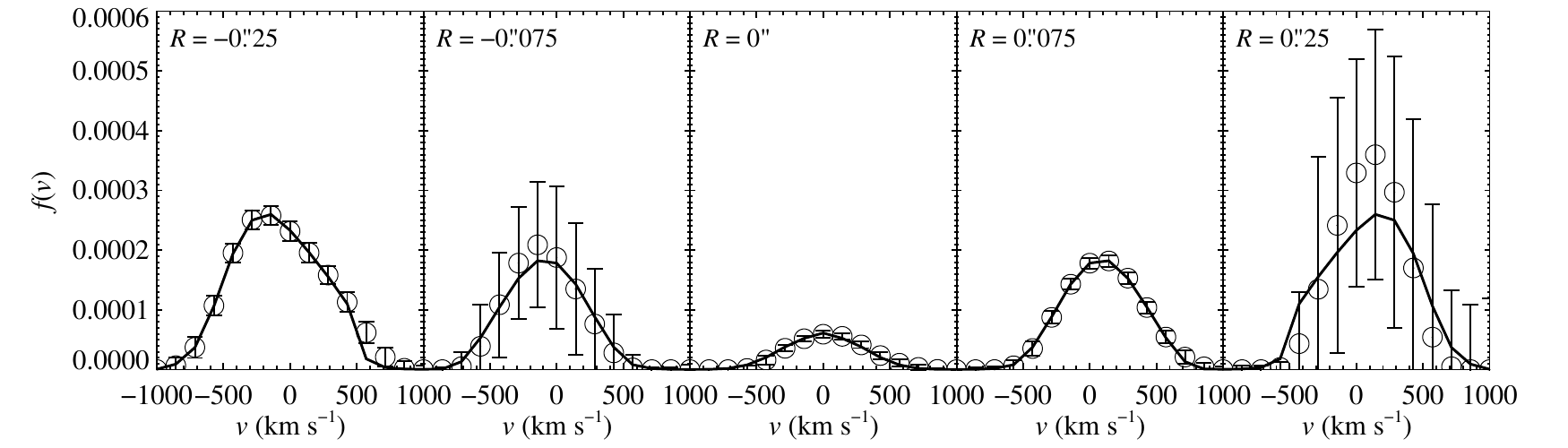}
\caption{LOSVDs for STIS data from the stellar ring. The circles with
  error bars are the LOSVD extractions from the STIS data, and the
  solid line is the best-fit model.  Each panel comes from a different
  spatial bin as indicated in Table \ref{t:n3706stisdata}.  The
  signal-to-noise ratio varies from spatial bin to spatial bin (as
  discussed in Section \ref{spectroscopy}), but is excellent in the
  first, third, and fourth bin. As the model is axisymmetric, the
  model LOSVDs at $R$ and $-R$ are mirror images.  The y-axis
  normalization is arbitrary but consistent across all panels of this
  figure and of Fig.\ \ref{f:losvd1} such that the integral of the
  curves is proportional the total light in each bin.  The differences
  in the integral of each panel are because of the different numbers
  of pixels in each radial bin, the different slit widths between this
  figure and Fig.\ \ref{f:losvd1}, and the different amount of
  light.}  \kgfigstarend{losvd0}{LOSVD of stellar ring from STIS
  data.}

\kgfigstarbeg{losvd1}
\includegraphics[width=\textwidth]{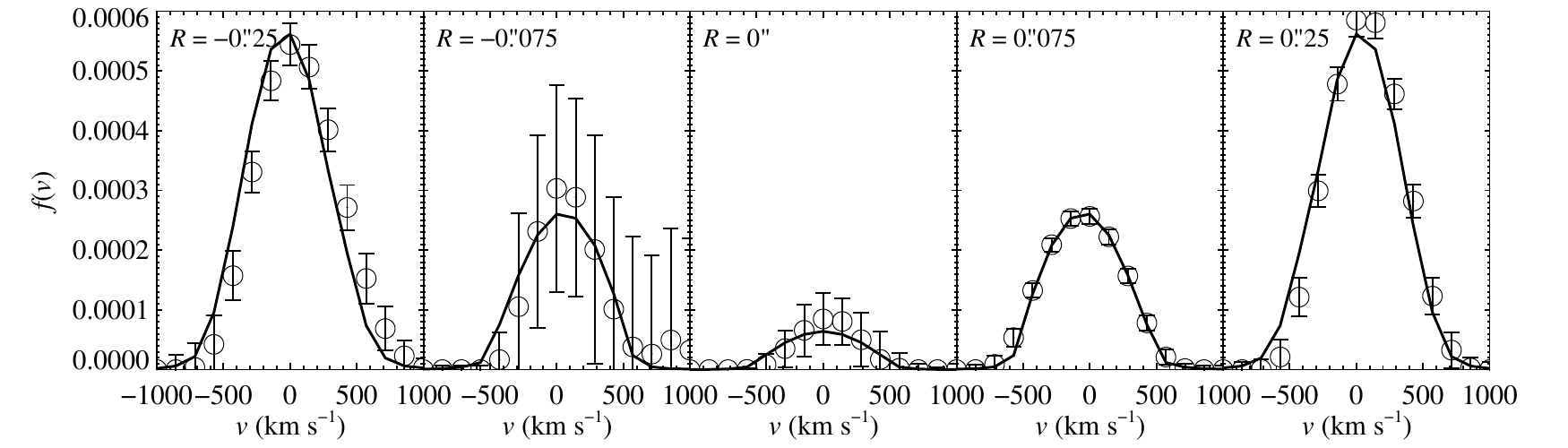}
\caption{Same as Fig.\ \ref{f:losvd0} but for the offset STIS slit position.}
\kgfigstarend{losvd1}{LOSVD of offset major axis from STIS data.}

The velocity profile (Fig.\ \ref{f:spec}) confirms that the stellar
feature is a ring rather than a filled disk as the limb brightening
indicated \citep{2002AJ....124.1975L}.  The velocity increases from
the center towards a maximum of $\sim100\ \kms$ at a projected radius
of $R = 0\farcs075$--$0\farcs1$.  If we had observed a disk without a
hole in its middle, the velocity dispersion would increase at $R
\approx 0$ (assuming there is a black hole at the center), where the
circular velocity is highest and the contributions from the two sides
of the disk are unresolved.  The velocity and dispersion profiles of
the STIS data are also plotted in Fig.\ \ref{f:compareparallel} as a
function of distance from the minor axis rather than distance from the
center (i.e., cylindrical radial coordinate rather than spherical
radial coordinate).  This allows one to compare the velocity profile
of the putative ring with that of a nearby but offset slit position.
Compared to the off-axis velocity profile, the on-axis profile shows
rapid rotation, as expected from a ring-like system, but with a large
velocity dispersion for its rotational speed ($|V|/\sigma \la 0.3$).

\kgfigstarbeg{spec}
\includegraphics[width=\textwidth]{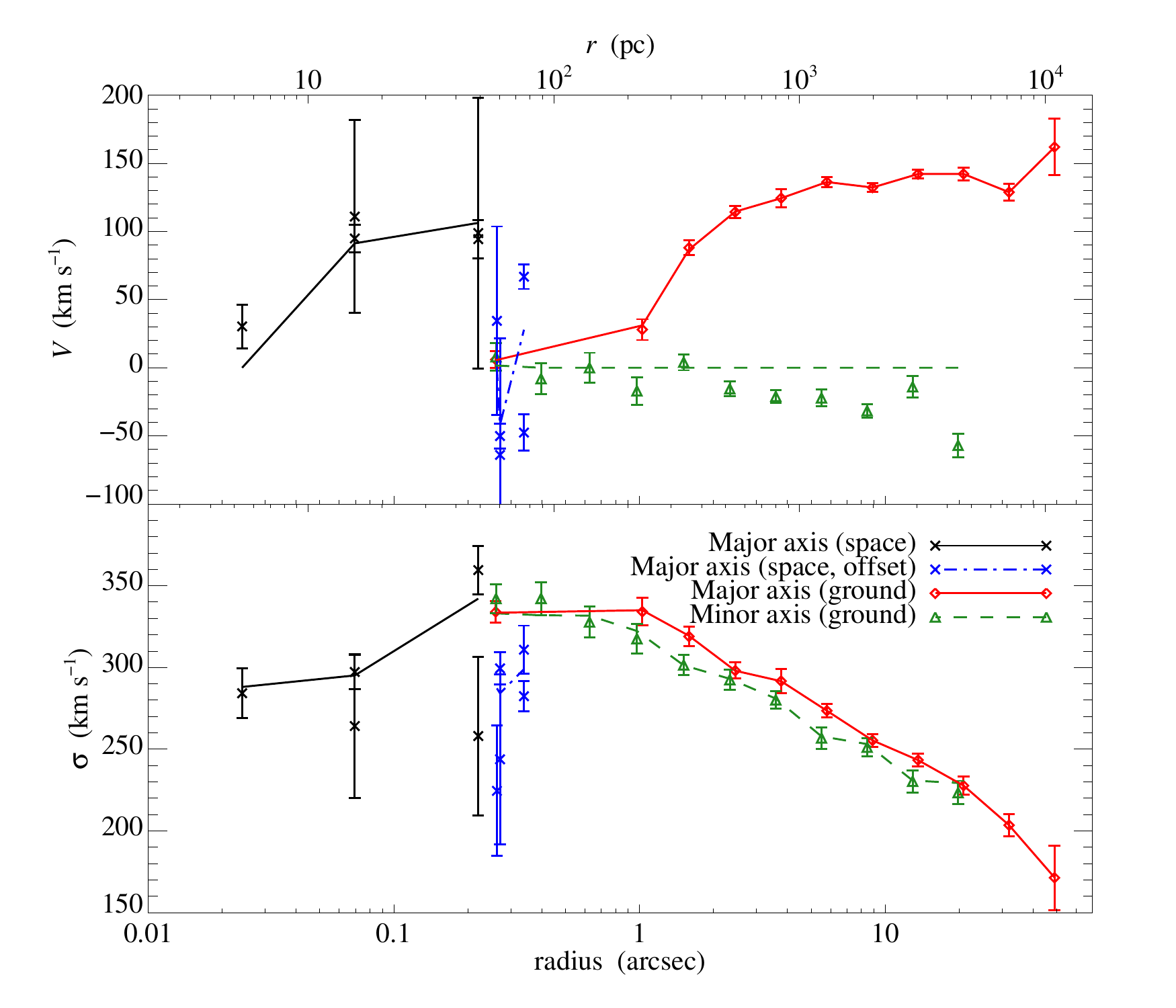}
\caption{Mean line-of-sight velocity and velocity dispersion of NGC
  3706 as a function of radius.  Symbols are the data, and lines are
  the from the best fit model (parameters: $M = 6.0 \times
  10^8\ \msun$, $\Upsilon_V = 6.0\ \msun\ \lsunv^{-1}$, $V_c =
  600\ \kms$, and $r_c = 30\ \units{kpc}$).  Black crosses are STIS
  major axis; blue crosses are STIS major axis offset by 0\farcs3
  perpendicular to the slit PA; red diamonds are ESO major axis (due
  to \citealt{1994MNRAS.270..523C}); and green triangles are ESO minor
  axis data.  Black solid is major axis; blue dot-dashed is offset
  major axis; red solid is major axis at ground resolution; and green
  dashed is minor axis at ground resolution.  The colors correspond to
  the slits in Figure \ref{f:slits}.  Data and models with $r = 0$ are
  plotted at $r = 0\farcs025$ on the logarithmic plot.  Note that for
  the STIS data, there are two data points for each radius $r >
  0\farcs025$ and slit position.  This is because velocity profiles
  from each side of the galaxy center were extracted independently.
  The model, however, is axisymmetric.  The ESO data from each side of
  the galaxy, by contrast, were averaged.}  \kgfigstarend{spec}{Mean
  line-of-sight velocity and velocity dispersion as a function of
  radius.}

\kgfigstarbeg{compareparallel}
\centering
\includegraphics[width=0.8\textwidth]{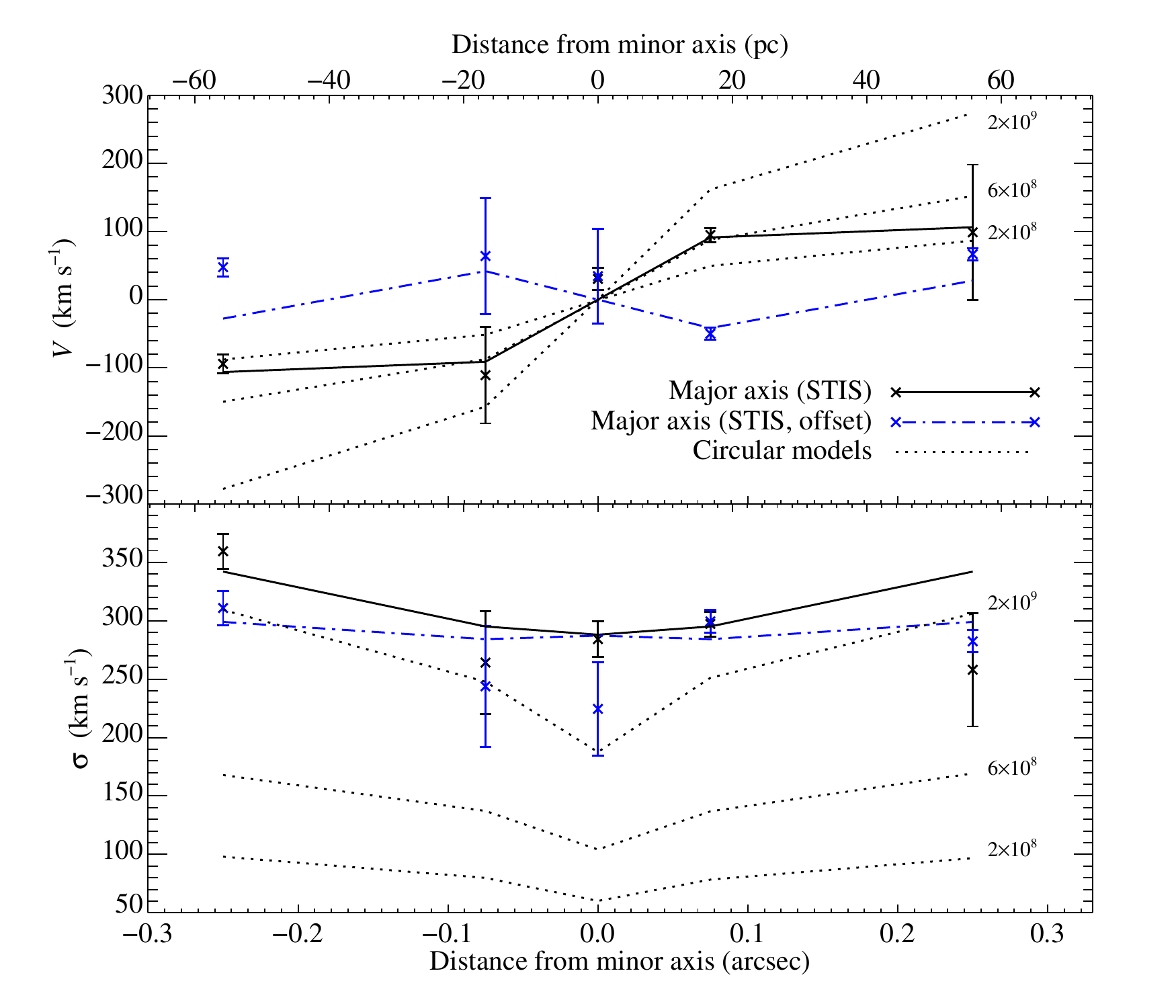}
\caption{Mean velocity and velocity dispersion profiles (symbols) with
best-fit models (lines) for two slit positions: on the major axis
(black symbols and black solid line) and parallel to the major axis
offset perpendicularly by 0\farcs3 (blue symbols and blue dot-dashed
line).  The abscissa is distance from minor axis, i.e., cylindrical
radial coordinate.  The much faster rotation speed seen in the on-axis
slit is consistent with a rapidly rotating, thin ring of stars.  The
dotted black lines are results from our simple circular-orbit models (Sect.\ \ref{kinematicsring})
with black hole masses indicated by the numbers at the right of each
line and should be compared to the on-axis data.  The low value of
$|V|/\sigma \la 0.3$ for the stellar ring cannot be explained with a
single population of co-rotating stars in circular orbits.}
\kgfigstarend{compareparallel}{Major axis and offset axis kinematics.}

\kgtabbeg{r@{\extracolsep{-1pt}}rr@{\extracolsep{0pt}$\pm$}lr@{$\pm$}lr@{$\pm$}lr@{$\pm$}l}
  \footnotesize
  \tablecaption{Kinematic Profile for NGC 3706 from STIS Observations}
  \tablewidth{\columnwidth}
  \tablehead{
     \colhead{$R$} &
     \colhead{Bin} &
     \multicolumn{2}{c}{$V$} &
     \multicolumn{2}{c}{$\sigma$} &
     \multicolumn{2}{c}{$h_3$} &
     \multicolumn{2}{c}{$h_4$} \\
     \colhead{(\arcsec)} &
     \colhead{(pix)} &
     \multicolumn{2}{c}{($\mathrm{km\ s^{-1}}$)} &
     \multicolumn{2}{c}{($\mathrm{km\ s^{-1}}$)} &
     \multicolumn{2}{c}{} &
     \multicolumn{2}{c}{} 
  }
  \startdata
$-$0.250 & 5 &  $-$94 &  14 & 360 &  15 & 0.087 & 0.005 & $-$0.060 & 0.015\\
$-$0.075 & 2 & $-$111 &  71 & 264 &  44 & 0.035 & 0.05 & $-$0.036 & 0.12\\
0.000 & 1 &  30 &  16 & 284 &  15 &  0.031 & 0.005 & $-$0.036 & 0.03\\
0.075 & 2 &  95 &  10 & 297 &  11 &  0.021 & 0.001 & $-$0.034 & 0.023\\
0.250 & 5 &  99 &  99 & 258 &  49 & $-$0.029 & 0.06 & $-$0.041 & 0.048\\
\hline\\[-1.5ex]
$-$0.250 & 5 & 48 &  13 & 311 &  15 & 0.052 & 0.009 & $-$0.025 & 0.03\\
$-$0.075 & 2 & 64 &  86 & 244 &  52 & 0.064 & 0.028 &  0.008 & 0.100\\
0.000 & 1 & 34 &  69 & 225 &  40 & $-$0.014 & 0.033 & $-$0.041 & 0.038\\
0.075 & 2 & $-$50 &  9 & 299 &  10 &  0.009 & 0.004 & $-$0.057 & 0.022\\
0.250 & 5 &  67 &  9 & 282 &  9 &  0.011 & 0.002 & $-$0.037 & 0.024
\enddata
\tablecomments{Gauss--Hermite moments for velocity profiles derived
from STIS data.  Radii are distances from the minor axis, i.e.,
cylindrical radial coordinate, given in arcsec with negative values indicating 
the southeast side and positive the northwest side of the slit.  The second column
gives the width of the radial bin in pixels, which are 0\farcs05.  The
top section gives values for the center positioning with the 0\farcs1
slit, and the bottom section gives values for the off-center
positioning with the 0\farcs2 slit.}
\kgtabend{n3706stisdata}{Gauss-Hermite moments of NGC 3706}

For wide-field kinematics ($R \gta 1\arcsec$), we use the velocity and
velocity dispersion profiles due to \citet{1994MNRAS.270..523C}, based
on \ion{Mg}{2} observations with the ESO 2.2-m telescope.  The data
were taken along the major and minor axes, \emph{defined by the isophotes at
large radii}.  The signal-to-noise ratio of these data was
limited, so that only the mean velocity and velocity dispersion---not
higher moments---were usable.  To work with our axisymmetric code, we
averaged the data from one side of the galaxy to the other.  
Despite the low signal-to-noise ratio of the data, these profiles are smooth
(Fig.\ \ref{f:spec}) with significant rotation and an increasing
dispersion from $\sim10\ \units{kpc}$ to $\sim200\ \units{pc}$.

From the ESO spectroscopy, we
measured an effective velocity dispersion of $\sigma_e = 325 \pm 5 \
\kms$, where 
\beq
\sigma^2_e \equiv \frac{\int_{0}^{R_e} \left[{\sigma(r)^2 + V(r)^2}\right] I\left(r\right) dr}{{\int_{0}^{R_e} I\left(r\right) dr}},
\label{e:sigmae}
\eeq
where $R_e$ is the effective radius, $I(r)$ is the surface brightness
profile, and $V(r)$ and $\sigma(r)$ are the mean velocity and velocity
dispersion of the LOSVD, and where the integral is along the major axis.  We use the \citet{laueretal05}
reported value of $R_e = 34\farcs9$ and integrate equation (\ref{e:sigmae}) with cubic spline interpolation and Newton--Cotes quadrature. Other interpolation schemes and quadrature techniques give similar results.

Others \citep[e.g.,][]{2012ApJ...756..179M} argue that the integral in equation (\ref{e:sigmae}) should start not at 0 but at the radius of the black hole's sphere of influence.  There are two potential definitions of the sphere
of influence.  The first is the classical definition: $R_\mathrm{infl}
= G M \sigma^{-2}$, where $\sigma$, which is generally a function of
radius, is evaluated at $R_\mathrm{infl}$, a recursive definition that
converges quickly.  The second is the radius at which the enclosed
stellar mass is equal to the black hole mass: $M_*(r<R_\mathrm{encl})
= M$.  Based on our results below, the classical sphere of influence for NGC 3706 is
$R_\mathrm{infl} = 25\ \units{pc} \approx 0\farcs11$, and the enclosed
stellar mass radius is $R_\mathrm{encl} = 34\ \units{pc} = 0\farcs15$.
Thus for both definitions, we are able to resolve the sphere of
influence of the black hole, a fact that can only be determined after
the black hole mass has been measured \citep{2011ApJ...738...17G}.  Using this alternative definition of $\sigma_e$ makes only a 1\% or 2\% difference.

\section{Modeling}
\label{modeling}

We use the \citet{1979ApJ...232..236S} orbit-based method to make
axisymmetric three-integral models of NGC 3706 using the
implementation described in \citet{gebhardtetal03} and
\citet{siopisetal08}.  For NGC 3706, the presence of the ring requires
an approach different from our usual one.  We perform a
photometric bulge--ring decomposition, deproject each component
separately, and then combine the deprojected luminosity densities into
a single luminosity density.

\subsection{Deprojection to luminosity density}
For the deprojection, we created a bulge surface brightness image
based on the surface brightness profile of the bulge's minor
axis.  To create the two-dimensional image, we used the apparent axis
ratio at large radii where there is no ring.  Inside 0\farcs5, the
presence of the ring makes determination of bulge axis ratio
impossible.  We extrapolated from $r = 0\farcs5$ to the center,
assuming no change in the bulge's axis ratio.  We then subtracted the
bulge-only image from the total image, leaving only the ring light.

Deprojection of the bulge light proceeded as usual under the
assumption that constant luminosity density contours are coaxial
spheroids.  Because the stellar ring is edge-on, we enforce an
inclination of $i = 90^\circ$.  This prescription results in a core
profile for the galaxy.

Deprojection of the ring light assumed that the ring is flat,
circular, and edge-on.  From the unresolved ring height 
($z < 0\farcs05$), the axis ratio at the inner edge is $z/r < 0.5$ 
and at the outer edge $z/r < 0.125$.   A complication in the 
modeling is that there is a misalignment of the galaxy bulge 
isophotes at large radii by about 42$^\circ$ from the ring major 
axis.  The ring major axis has $\mathrm{PA} = 115^\circ$ at radii 
$r \le 0\farcs4$, whereas the bulge has $\mathrm{PA} = 73^\circ$ 
at radii $r > 6\arcsec$ (Fig.\ \ref{f:slits}).  The change in PA, 
or isophotal twist, occurs sharply between semimajor axes of 1 and 
2\arcsec\ \citep{2002AJ....124.1975L}.  In our model we simply 
combine the luminosity densities of the two components as if 
they were coaxial.

A frequent cause of isophotal
twists is the projection of a triaxial light distribution, but in this
case it is more likely that the change in PA represents the transition
from an outer axisymmetric light distribution to an inner
distribution, dominated by the ring light, that is axisymmetric
relative to a different axis.  

\subsection{Orientation of data and model}

The ground-based spectra come from slits positioned on the galaxy's 
major and minor axes at large radii, as is appropriate since these 
data are used to constrain the large-scale kinematics.  In our modeling of
the galaxy, we calculate LOSVDs at the central bins of the
ground-based slits as if the ring were coaxial with the galaxy bulge.
This cannot have any effect because any differences between our model
and how the galaxy actually appears are beneath the resolution limit
of the ground spectra.

It is possible that the misalignment between our assumed axis of
symmetry and the actual axis of symmetry of the outer regions of the
galaxy could affect our inference of the stellar mass-to-light ratio.
Because of the covariance between black hole mass and stellar
mass-to-light-ratio, this could result in incorrect inferences of
black hole mass.  This also is unlikely to have made a large
difference.  The mass-to-light ratio that we find best describes the
data under our assumptions is entirely reasonable for a galaxy of this
size and close to the value obtained by \citet{1994MNRAS.270..523C}.
Finally, if the stellar ring is kinematically decoupled from the rest
of the galaxy, as seems reasonable, then modeling the galaxy as having
a coaxial stellar ring should not produce substantially different
kinematics.  The known deviations from our assumptions about the light
distribution, however, are unquantifiable systematic errors that are
not reflected in our quoted uncertainties.

\subsection{Model parameters}
Assuming constant but unknown stellar mass-to-light ratio
$\Upsilon_V$, and assuming that the ring and the rest of the galaxy
have the same mass-to-light ratio, we generate a stellar mass density,
$\rho(r, \theta)$.  To this we add a point mass of unknown mass,
$\mbh$, at the center of the galaxy and a dark matter halo potential,
parametrized as a spherical cored logarithmic halo with mass density profile
\beq
\rho(r) = \frac{V_c^2}{4\pi G} \frac{3 r_c^2 + r^2}{(r_c^2 + r^2)^2},
\label{coredhalo}
\eeq
with unknown asymptotic circular velocity, $V_c$, and core radius,
$r_c$.  Each unknown, $\Upsilon_V$, $\mbh$, $V_c$, and $r_c$ becomes a
parameter of our model.  We calculate orbits of representative stars
in the combined potential of the stars, black hole, and dark matter
halo, keeping track of the orbit's contribution to the galaxy's
surface brightness and line-of-sight-velocity distributions (LOSVDs)
in each bin where we have data.  The number of orbits is different for
each set of parameters but ranged from $\sim12,000$--$17,000$ with
$\sim15,000$ a typical value.  We calculate the set of non-negative
weights of the orbits that minimizes the difference between the LOSVDs 
of the model and the data, subject to the constraint of matching the surface 
brightness throughout the galaxy.  This brief description of our kinematic modeling
necessarily simplifies many of the details, which are expanded in
\citet{gebhardtetal03}, \citet{2000AJ....119.1157G},
\citet{2004astro.ph..3257R}, \citet{2004MNRAS.353..391T},
\citet{2005MNRAS.360.1355T}, and \citet{siopisetal08}. Similar models
are described by others \citep{1984ApJ...286...27R,
1997ApJ...488..702R, 1999ApJS..124..383C, 2004ApJ...602...66V,
vandenboschetal08}.

To determine what range of parameters to use, we ran an initial coarse
grid in the 4-dimensional parameter space and then used a refined,
uniform grid around the best models from the coarse grid.  The final
range in parameters was $\Upsilon_V = 3$--$9\ {\msun\ \lsunv^{-1}}$;
$\mbh = 0$--$2.2 \times 10^9\ {\msun}$; $V_c = 100$--$600\ \kms$; and
$r_c = 5$--$50\ \units{kpc}$.  We also ran models with no dark matter
halo, parametrized as $V_c = 0.01\ \kms$ and $r_c = 1000\ \units{kpc}$.
The gridding of the parameters can be ascertained from Figure
\ref{f:rcchisq}.

\subsection{Modeling results}
The best-fit model has parameters $M = 6 \times 10^8\ \msun$,
$\Upsilon_V = 6.0\ {\msun\ \lsunv^{-1}}$, $V_c = 600\ \kms$, and $r_c =
30\ \units{kpc}$.  Models with $M = 0$ were ruled out at very high
significance, $\Delta\chi^2 = 15.4$.  Marginalizing over the other
parameters, the 1, 2, and 3$\sigma$ uncertainty ranges for $M$ are
$5.1$--$6.7 \times 10^{8}$, $3.2$--$8.9 \times 10^{8}$, and $1.4$--$12
\times 10^{8}\ \msun$, respectively.  The corresponding uncertainty
ranges for $\Upsilon_V$ are $5.8$--$6.2$, $5.3$--$6.6$, and
$4.7$--$7.0\ {\msun\ \lsunv^{-1}}$, respectively.  Models without a dark
matter halo are ruled out at a significance of $\Delta\chi^2 > 64$,
but we cannot disentangle the strong correlation between $r_c$ and
$V_c$ to constrain these parameters individually.  This is to be
expected without good coverage at large radii
\citep[e.g.,][]{2011ApJ...729..129M}.  We can, however, put weak
limits on the central density, i.e., $\rho_c \equiv 3 V_c^2 / 4 \pi G
r_c^2$.  The best fit value is $\rho_c = 0.02\ \msun\ \units{pc^{-3}}$
with 1, 2, and 3$\sigma$ uncertainty ranges of $0.016$--$0.04$,
$0.005$--$0.12$, and $0.004$--$0.4\ \msun\ \units{pc^{-3}}$.  
It is worth noting that our best-fit value for $V_c$ 
is at the maximum value considered for this parameter.  Given that there is no 
sign of convergence, that there is a degeneracy between $V_c$ and $r_c$, 
and that we do not have strong data at large data where we would best be 
able to get information about the dark matter halo, we did not spend more 
computational time to find the global minimum of $V_c$.  Our results suggest 
that as long as some plausible dark matter halo  is included in our 
models, our inferences about the black hole mass and stellar 
mass-to-light velocity are not strongly affected, even if we cannot 
constrain the parameters of the dark matter halo.

The projection of $\chi^2$ versus $M$ and versus $\Upsilon$ (Fig.\
\ref{f:rcchisq}) reveals the importance of including a dark matter
halo in the mass model.  First, we note that for NGC 3706, the models
without a dark matter halo (shown as red diamonds in the figure) still
rule out the absence of a black hole at high significance, and the
mass is consistent with the dark matter halo models at about the
$2.5\sigma$ level.  The best fitting models without a dark halo have
higher $\Upsilon_V$ values in order to account for the larger amount
of mass at large radii.  Since there is a small covariance between
$\Upsilon_V$ and $M$, this results in a smaller black hole mass.  The
difference between the inferred black hole masses when including and
not including a dark matter halo are smaller when the black hole's
gravitational sphere of influence is well resolved, a condition that
minimizes the degeneracy \citep{2009ApJ...700.1690G,
2011ApJ...729...21S, gebhardtetalm87, 2013arXiv1306.1124R}.

\kgfigstarbeg{rcchisq}
\includegraphics[width=0.45\textwidth]{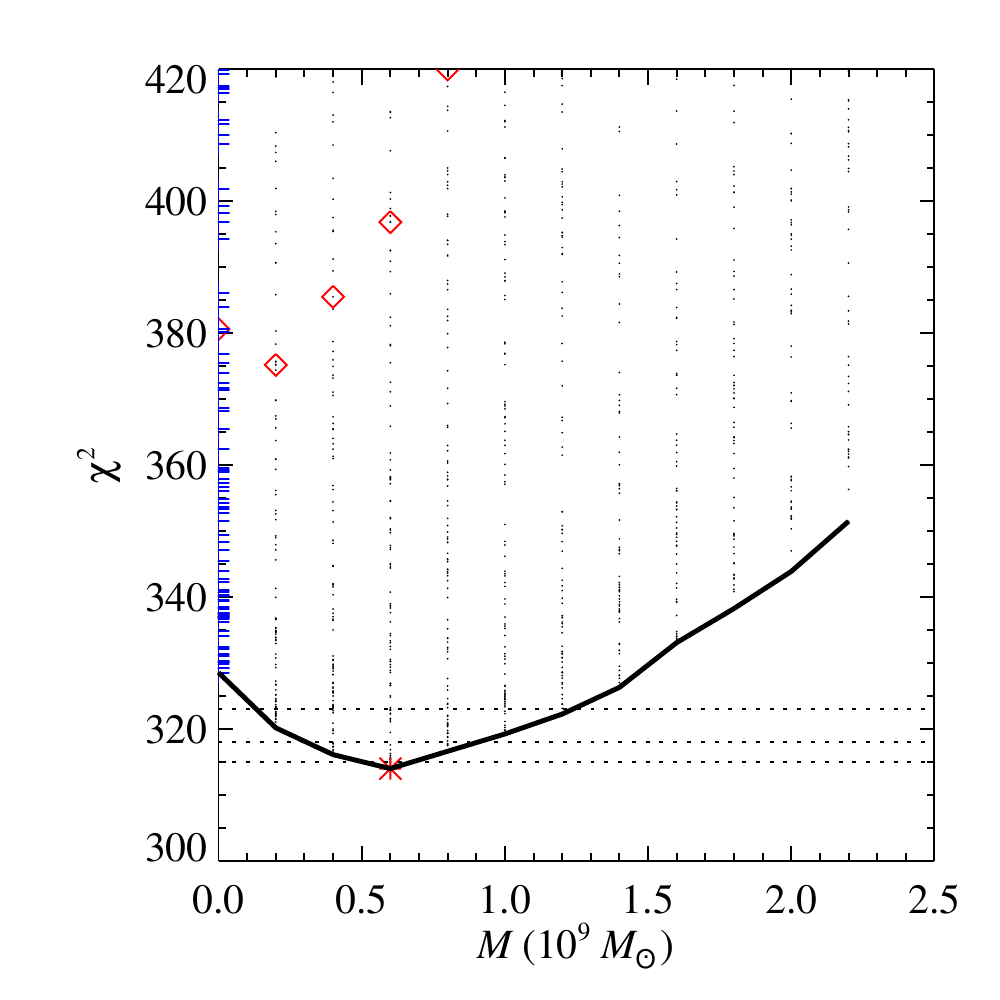}
\includegraphics[width=0.45\textwidth]{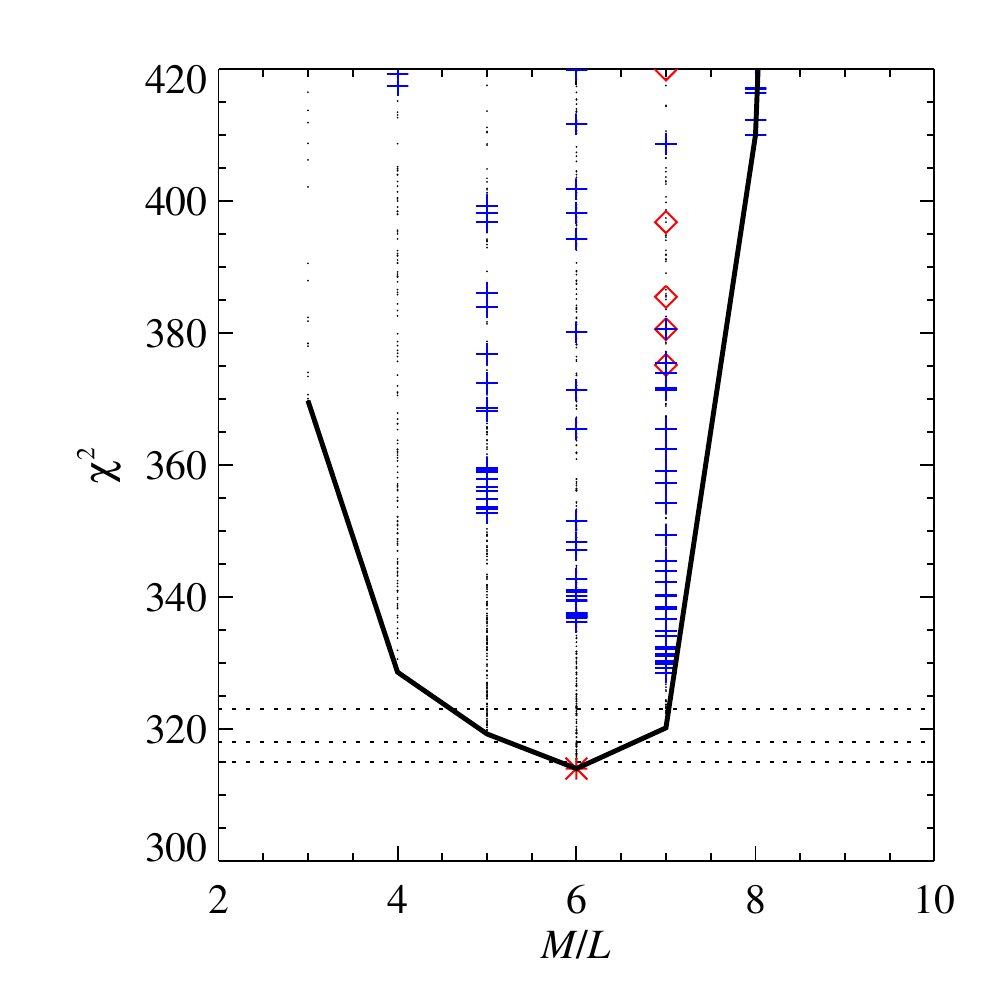}\\
\includegraphics[width=0.45\textwidth]{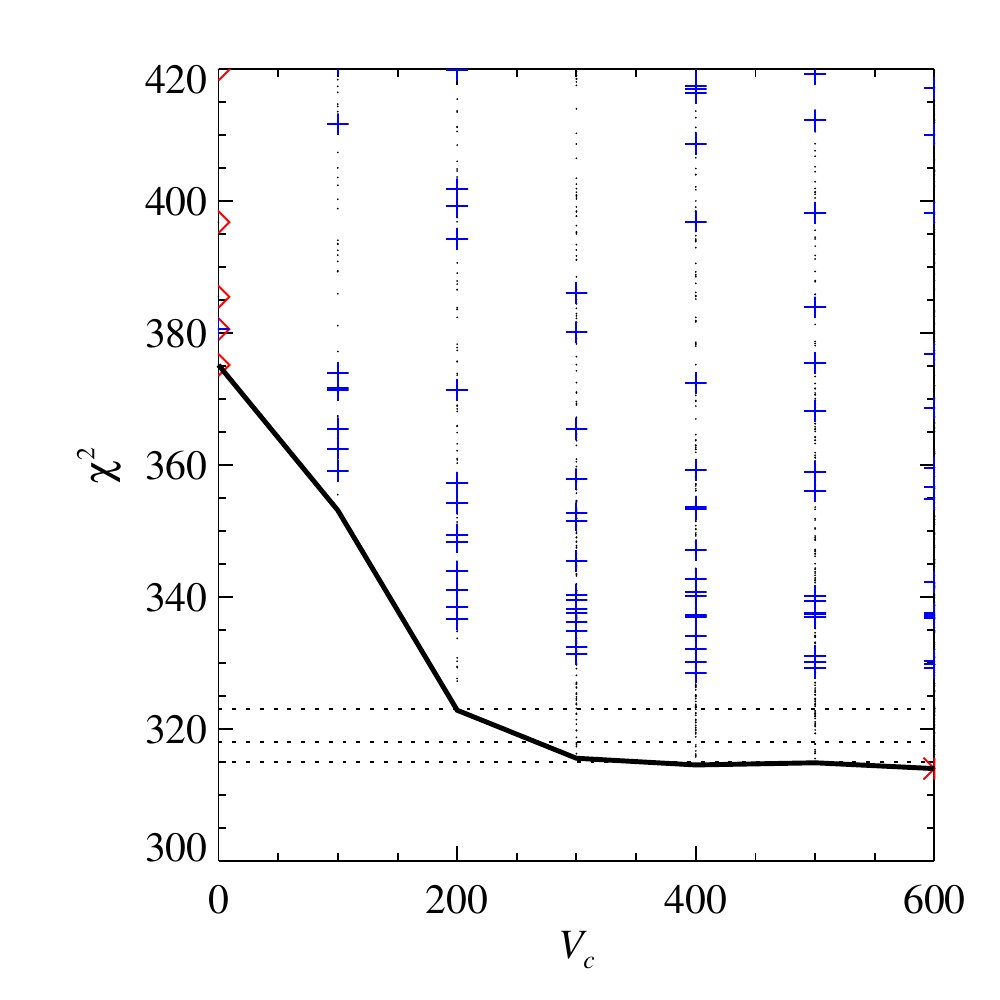}
\includegraphics[width=0.45\textwidth]{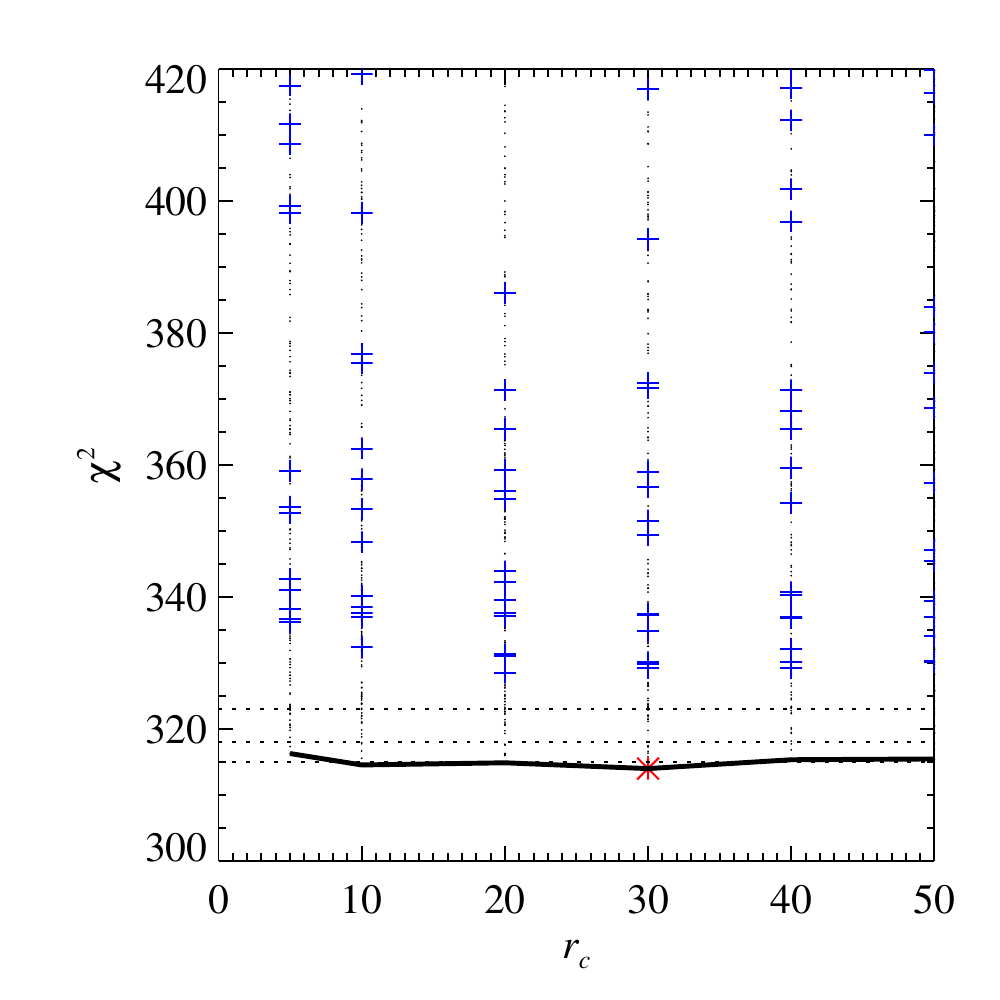}
\caption{Best-fit models. Each panel plots $\chi^2$ for all
models as a function of the given parameter ($M$, $\Upsilon_V$, $V_c$, and
$r_c$), projecting through the other three dimensions.  All models are
plotted as black dots.  The best-fit model ($M = 6.0 \times 10^8\
\msun$, $\Upsilon_V = 6.0$, $V_c = 600\ \kms$, and $r_c = 30\
\units{kpc}$) is shown with a red asterisk.  Models with $M = 0$ are
over-plotted with blue crosses, and models with no dark matter halo
($V_c = 0.01\ \kms$ and $r_c = 1000\ \units{kpc}$) are over-plotted with
red diamonds.  Each panel has three dotted lines that show, from
bottom to top, $\Delta\chi^2$ values above the minimum of 1, 4, and 9,
corresponding to 1, 2, and 3$\sigma$ confidence intervals of each
individual parameter.  A heavy black line is drawn connecting the smallest
$\chi^2$ values for each parameter value.  We have good constraints on
black hole mass $M = 6 \times 10^8\ \msun$ with 1, 2, and 3$\sigma$
uncertainty ranges of $5.1$--$6.7\ \times 10^{8}$, $3.2$--$8.9\ \times
10^{8}$, and $1.4$--$12\ \times 10^{8}\ \msun$, respectively.}
\kgfigstarend{rcchisq}{Modeling results / chi-square curves.}

\section{Discussion and Conclusions}
\label{discussion}

\subsection{The mass of the black hole}
The mass of the black hole can be predicted from the velocity
dispersion of its host galaxy \citep[the \msigma\ 
relation;][]{gebhardtetal00a, fm00, tremaineetal02,   
2009ApJ...698..198G, 2013ApJ...764..184M, 2013ARA&A..51..511K} or
its host (classical) bulge luminosity \citep[the \ml\ 
relation;][]{kormendy93a, kr95, magorrianetal98, kg01,
2004ApJ...604L..89H, 2009ApJ...698..198G, 2011Natur.469..374K,
2011MNRAS.413.1479S, 2013ApJ...764..184M, 2013ARA&A..51..511K}.

The effective velocity dispersion we measured in Section
\ref{spectroscopy} is $\sigma_e = 325 \kms$.  The \msigma\ relation
for ellipticals and classical bulges due to
\citet{2013ARA&A..51..511K} predicts a black hole mass and 68\%
distribution interval of $\mbh = 2.6_{-1.3}^{+2.5} \times
10^{9}\ \msun$.  Our best fit mass estimate for the black hole in NGC
3706 is 4.3 times smaller than the \msigma\ relation predicts, or
about a $2.2\sigma$ outlier compared to the 0.29 dex rms intrinsic
scatter.  If we exclude the inner 0.4 or 1\arcsec\ from the integrals
in Equation (\ref{e:sigmae}) to avoid possible contamination by the
ring, then $\sigma_e$ reduces to 318 or $300\ \kms$, respectively,
which would put NGC 3706 closer to the ridge line relation, enough to
make it only a $1.7\sigma$ low outlier.  A dispersion of $300\ \kms$
also brings makes it less of an outlier from the \citep[i.e.,
  $L$--$\sigma$]{1976ApJ...204..668F} relation for either a core or
coreless elliptical \citep{2013ApJ...769L...5K}.

The bulge absolute magnitude of NGC 3706 is $M_V = -22.26$.  A typical
$V - K$ color for a galaxy of NGC 3706's luminosity is $V - K = +2.98$
\citep{2013ARA&A..51..511K} so that $M_K = -25.24$.  At this bulge
luminosity, the \citet{2013ARA&A..51..511K} \ml\ relation for
ellipticals and classical bulges predicts $\mbh = 1.7_{-0.9}^{+1.7}
\times 10^{9}\ \msun$.  The black hole in NGC 3706 is about 2.9 times
smaller, or a $1.5\sigma$ outlier for the 0.30 dex rms scatter.

The scaling relations of \citet{2013ARA&A..51..511K} make larger
predictions for black hole mass than earlier scaling relations
\citep[e.g.,][]{tremaineetal02, 2009ApJ...698..198G}.  The difference
in predictions arises mostly from the different intercepts of the
relations; the slopes are very similar.  The primary reasons for the
difference in intercepts are the upward revisions in black hole masses
for several sources, the larger number of core ellipticals with large
black hole masses, and the rejection of some black hole mass
measurements from emission-line rotation curve techniques of poor
quality.  Other differences are that \citet{2013ARA&A..51..511K} did
not include galaxies with only upper limits to the black hole mass,
and excluded pseudobulges, late-type galaxies, ``monster black
holes,'' and recent mergers, on the grounds that the galaxies are
physically very different from the galaxies comprising the bulk of the
sample.  If, despite the strong arguments in
\citet{2013ARA&A..51..511K}, we include the pseudobulges, late-type
galaxies, ``monster black holes,'' and recent mergers and re-fit the
\msigma\ and \ml\ relations, the new fits to the resulting larger,
more diverse sample do not significantly alter the slope or intercept,
but they do increase the intrinsic scatter to 0.46 and 0.68 dex for
the \msigma\ and \ml\ relations, respectively.  NGC 3706 is still
below these forms of the relations, but only by 1.4 and 0.7$\sigma$
for the \msigma\ and \ml\ relations, respectively.

NGC 3706's diffuse starlight extending beyond the well defined bulge,
which may be Malin shells, and nuclear stellar ring are both
plausibly---but not conclusively---attributed to merger activity.  The
stellar ring could last for a Hubble time and therefore be a remnant
of an earlier merger before a more recent event that resulted in the
Malin shells.  Several other galaxies that are classified as ``mergers
in progress'' by \citet{2013ARA&A..51..511K} also fall below (or to
the right of) the scaling relations, especially the \ml\ relation.
Thus, it is not without precedent to find such a small black hole in a
host galaxy of this size.  Without deep images, however, it is
difficult to be certain whether NGC 3706 is a ``merger in progress'' as
defined by \citet{2013ARA&A..51..511K}.  NGC 3706 does differ from
other mergers in progress in that it is just as much an outlier in the
\msigma\ relation as it is in the \ml\ relation, whereas the other
mergers in progress are closer to the \msigma\ ridge line than they
are to the \ml\ ridge line.  Finally we note that the case for NGC
3706 as a merger in progress is far more circumstantial than for the
other galaxies identified as such in \citet{2013ARA&A..51..511K} and
that any merging activity in NGC 3706 could have happened as long as
$10^{9}\ \units{yr}$ ago.  In any case, as it is only a $\sim2\sigma$ outlier
from the scaling relations, the deviations may not be physically
meaningful.

\subsection{The kinematics of the ring}
\label{kinematicsring}
We did not explicitly constrain our model to use a ring, only that it
reproduce (1) the deprojected light distribution, which includes what
appears to be a stellar ring, and (2) the observed line-of-sight
velocities.  Nonetheless, the best orbit models were those that
included orbits constituting a stellar ring.  Fig.\ \ref{f:aniso}
plots the kinematics of the best fitting model in the equatorial
plane.  At the range of radii marked by shading in the figure
($0\farcs1 \la r \la 0\farcs4$), the ring should dominate the
light-weighted kinematics.  In this range there is a marked decrease
in dispersion in the radial and altitude angle directions (drops in
$\sigma_r$ and $\sigma_\theta$).  This is exactly as expected for a
stellar ring.

\kgfigstarbeg{aniso}
\centering
\includegraphics[width=\columnwidth,angle=270]{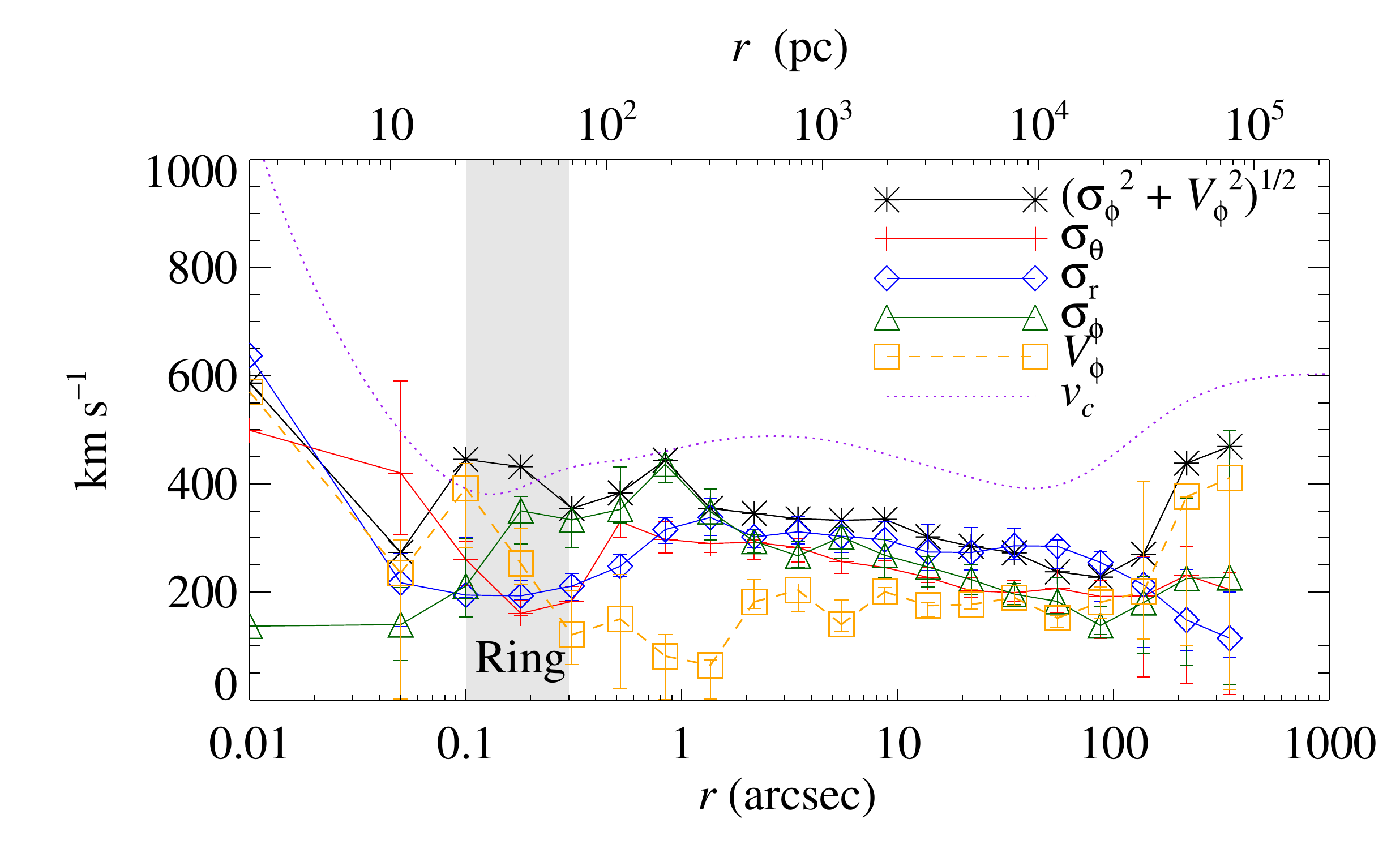}
\caption{ Kinematics of the best fitting orbit model in the equatorial
  plane.  Each symbol shows the dispersion in the direction indicated
  by the legend, where $r$ is the radial direction, $\phi$ is the
  tangential direction, and $\theta$ is the altitude angle (i.e.,
  angle above the equatorial plane) direction.  In the equatorial
  plane $\sigma_\theta = \sigma_z$.  The error bars  show the
  ranges of values derived from models within $\Delta\chi^2 < 4.7$
  ($1\sigma$ confidence for 4 parameters) of the minimum.  The region
  roughly occupied by the stellar ring is shaded light gray.  In the
  region of the ring, $\sigma_\theta$ and $\sigma_r$ decrease markedly
  while $\sigma_\phi$ and $(\sigma_\phi^2 + V_\phi^2)^{1/2}$ are
  steady or increase within the ring region.  The decrease in
  $\sigma_\theta$ and $\sigma_r$ is as expected by a stellar ring.  We
  also show the circular velocity calculated from the potential of the
  best-fitting model ($v_c^2 = r \nabla \Phi$).}
\kgfigstarend{aniso}{Best-fit model kinematics.}

We also fitted the major-axis STIS kinematics with a simple model in
which the ring is composed of stars on circular orbits, all traveling
in the same direction, conforming to the observed surface brightness
in the potential of a central point mass and a reasonable
``background'' potential.  The low value of $|V|/\sigma$ in the ring,
however, is inconsistent with such models.  The rotation profile can
be matched with a central mass much smaller than found by our
Schwarzschild modeling, and the dispersion profile can be matched by a
much larger mass, but simple circular stellar orbits could not
reproduce all of the kinematic data (Fig.\ \ref{f:compareparallel}).
Two potential explanations are that (1) the stellar orbits in the ring
are non-circular or (2) a large fraction of the stars in the ring are
counter-rotating.  Non-circular orbits would increase the
line-of-sight dispersion.  To consider this possibility, we calculated
the line-of-sight velocity and velocity dispersion as a function of
radius for stars in a spherical potential that has the same enclosed
mass profile as the non-spherical potential of NGC 3706.  The orbits
considered were rosette orbits with apocenters and pericenters
corresponding to the outer and inner extents of the stellar ring.
These rosette orbits were able to reproduce the high observed velocity
dispersion, but not the low observed $|V|/\sigma$.  Thus, we conclude
that the stars in the stellar ring are orbiting in both senses.

If a significant fraction of the stars are counter-rotating, then the
mean velocity of stars at a given radius would tend closer to zero and
the dispersion would increase.  This is the solution found by our best
fitting model with approximately $1/3$ of the stars in the ring
counter-rotating.  A partially counter-rotating ring sufficient to
drop the line-of-sight $|V|/\sigma$ to the observed ratio necessarily
implies LOSVDs that are asymmetric about their mean in the ring.  This
asymmetry is seen in the data as can be ascertained from the high
value of the third Gauss-Hermite moment of $h_3 = 0.09$ at $R =
-0\farcs25$ and the only negative value at $R = +0\farcs25$ of $h_3 =
-0.029$ (Table \ref{t:n3706stisdata}).  The LOSVD of the $R =
-0\farcs25$ observational bin can be seen to be asymmetric in the
right panel of Figure\ \ref{f:ringlosvd}.  This observed LOSVD is well
matched by the best-fit model (black solid line in
Fig.\ \ref{f:ringlosvd}).  The model LOSVD is composed mostly of (i) a
bimodal distribution of circular and nearly circular orbits in both
directions (red crosses), (ii) a unimodal distribution of more
eccentric orbits that orbit in both directions (blue diamonds), and
(iii) orbits of stars outside of the ring seen in projection at the
location of the ring (brown $\times$ symbols).  The distinction
between the first two categories is subtle and depends on whether the
orbit is entirely in the region where the ring is highly visible, as
it is for (i), or if, as it is for (ii), only the pericenters of the
orbits peak into the ring while the apocenter is outside, in the
region where the light of the ring blends into the background light of
the galaxy.  The orbit library in our best-fit model contains 31,804
orbits, of which 168 are type (i).  Of those 168 orbits, 13 have non-negligible
weights in our orbit reconstruction and make up the system in our
models (Fig.\ \ref{f:ringlosvd}, right panel, thick red line).  Thus
the solution found by the model includes a stellar ring with stars
orbiting in both directions.

\kgfigstarbeg{ringlosvd}
\centering
\includegraphics[height=0.4\textwidth]{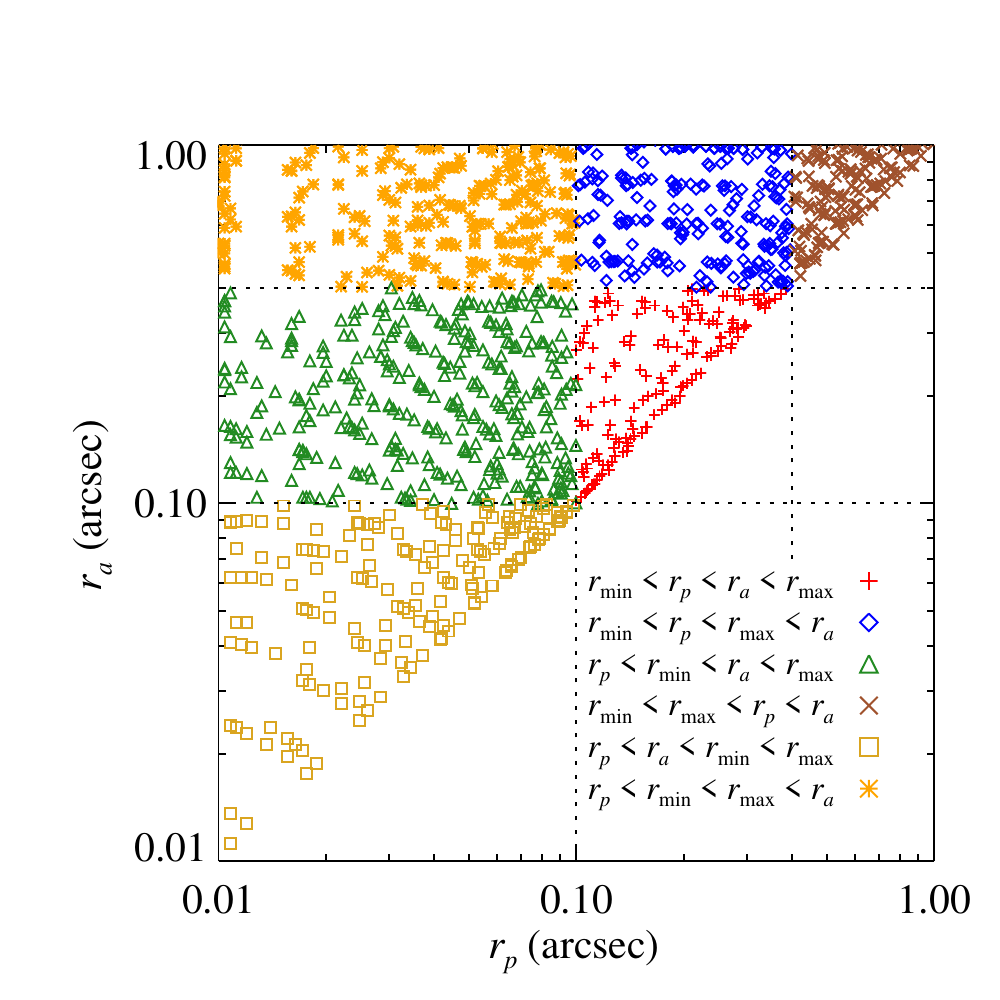}
\includegraphics[height=0.37\textwidth]{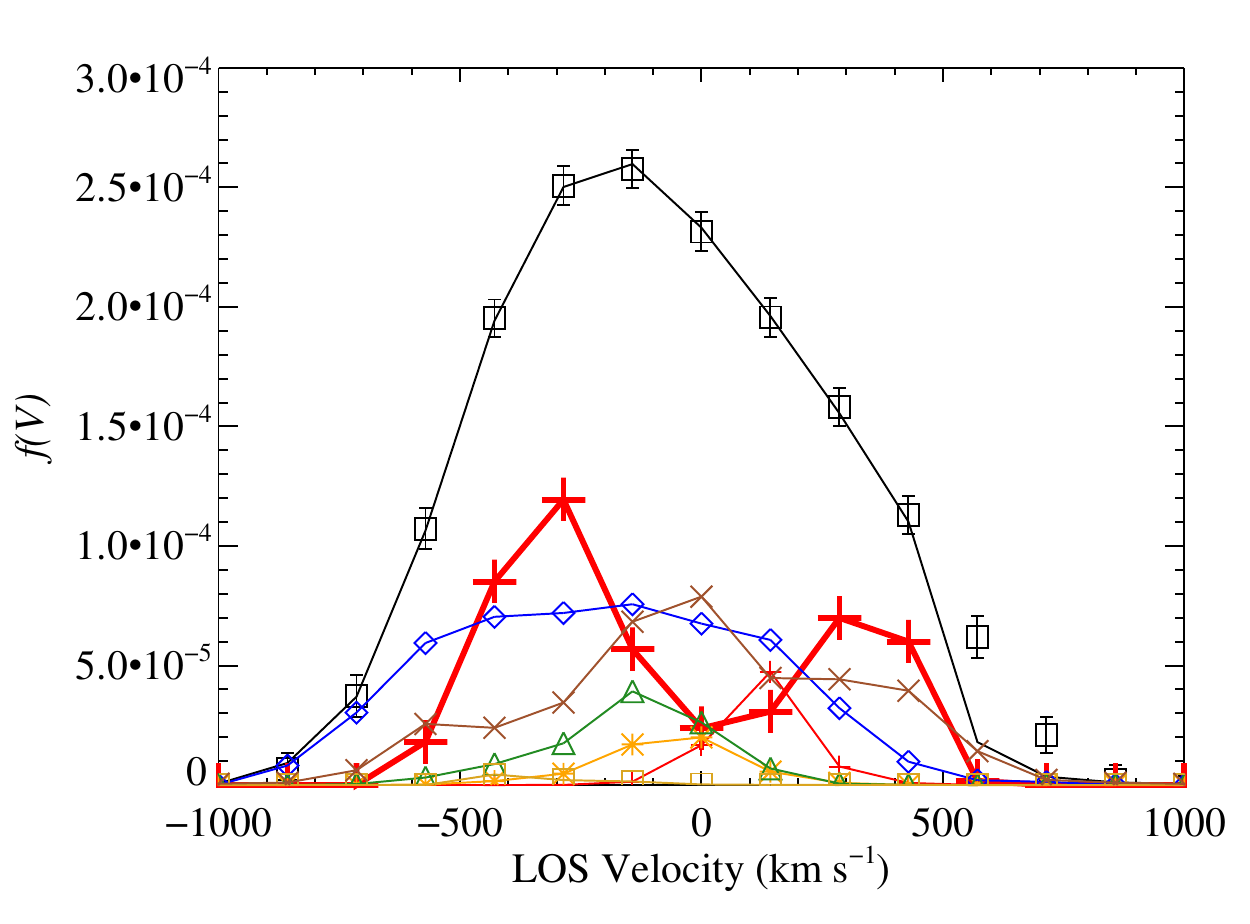}
\caption{ The left panel shows a portion of the
pericenter-distance--apocenter-distance ($r_p$--$r_a$) plane and
explains how we categorize orbits in the right panel depending on the
relation of the orbits' $r_p$ and $r_a$ compared to the ring minimum and
maximum radii, $r_\mathrm{min} = 0\farcs1$ and $r_\mathrm{max} =
0\farcs4$, shown in dotted lines.  Red $+$ symbols are orbits that are
entirely inside the ring's radial extent.  Blue $\Diamond$ symbols
have $r_p$ that lies within the ring's radial extent but $r_a$ that is
external to the ring.  Green $\bigtriangleup$ symbols have $r_a$ that
lies within the ring's radial extent but $r_p$ that is internal to the
ring.  Brown $\times$ symbols are entirely external to the ring.
Goldenrod $\Box$ symbols are entirely interior to the ring.  Finally
orange $\ast$ symbols have $r_p$ interior to the ring and $r_a$
exterior to the ring.
The right panel plots the line-of-sight velocity distribution (LOSVD)
of stars in the vicinity of the greatest extent of the stellar ring.
The spatial extent is $0\farcs125 < r < 0\farcs375$ and $|z| <
0\farcs05$.  The black $\Box$ symbols with error bars are the STIS
data.  The black line is the total best-fit model LOSVD and generally
closely matches the observations except for underpredicting the number
of large positive velocity orbits.  The colored symbols with lines
connecting them are contributions to the total model LOSVD from orbits
as categorized in the left panel.  LOSVD arising from orbits with
$r_\mathrm{min} < r_p < r_a < r_\mathrm{max}$ (red $+$ symbols) are
plotted separately for orbits that are within the plane of the ring
(thicker line) and for orbits that come out of the plane (thinner line).
  The total LOSVD (data and model) shows obvious asymmetric deviations
from a pure Gaussian, as expected from consideration of its large
third Gauss-Hermite moment, $h_3 = 0.09$.  The low value of
$|V|/\sigma$ could not be modeled by a sum of co-rotating circular
orbits.  The red $+$ symbols show that the orbits entirely within the
radial extent of the ring are bimodal in velocity distribution, and
the blue $\Diamond$ symbols show that the orbits with $r_p$ within
most of the ring but $r_a$ immediately outside of the ring have a
broad, unimodal distribution with a significant fraction orbiting in
the opposite sense of the median of distribution.  Thus the best-fit
model includes a ring with stars rotating in both senses.  }
\kgfigstarend{ringlosvd}{Asymmetric LOSVDs of the stellar ring.}

A potential concern is that since we can only observe projected
velocity distribution along the line of sight, what appears to be a
ring that has stars rotating in both directions could actually be a
ring that rotates entirely in one direction but in the opposite sense
of the rest of the galaxy.  First, we note that a
mono-counter-rotating ring can be calculated by our axisymmetric
models but was not the best-fit solution.  To directly address this
concern, we performed a simple kinematic decomposition of the ring
from the background galaxy (Fig.\ \ref{f:subtractedlosvds}).  Using
only the STIS data, we took on-axis LOSVDs from the location of the
ring and subtracted a weighted offset LOSVD at the same distance from
the minor axis.  This decomposition makes the reasonable assumption
that the bulge LOSVD just off the major axis is similar to the bulge
contribution to the LOSVD on the major axis.  The weight should be
equal to the fraction of light in the on-axis bin that does not come
from the ring.  That is, it is foreground and background light due to
the bulge of the galaxy.  Based on extrapolation of the surface
brightness profile, roughly $1/3$ of the light could come from stars
not in the ring.  The resulting decomposed ring LOSVDs, however, still
have a substantial component that rotates in the opposite sense to the
rest of the ring (Fig.\ \ref{f:subtractedlosvds}). Even if we assume
that half of the light in the vicinity of the ring actually comes from
foreground and background stars---and ignoring the unphysically
negative values for the LOSVD that result---there is still a
substantial counter-rotating component to the ring (i.e., stars with
negative velocity in the left panel of Figure 7, and with positive
velocity in the right panel).  The effect is more pronounced at the
ansae, (Fig.\ \ref{f:subtractedlosvds}; $R = -0\farcs250$) where the
differences in line-of-sight velocity of the counter-rotating
components are maximized.  So it is not possible to self-consistently
model the ring data without a counter-rotating component.

We note that the kinematics of the central stellar feature, in
particular the low $|V|/\sigma$, can be produced by an edge-on bar
structure.  The photometry, however, is inconsistent with a bar.
Galaxy-scale bars tend to be brightest at the center with either an
exponential or a slow, nearly constant decrease in surface brightness
along the bar major axis \citep{1985ApJ...288..438E,
  1989AJ.....98.1588K, 1993RPPh...56..173S}; and nuclear bars, though
less well cataloged, perhaps decrease very steeply
\citep[e.g.,][]{2001A&A...379L..44A}.  The central local minimum in
surface brightness is at odds with a filled-out bar structure and is
more naturally explained by a ring.

\kgfigstarbeg{subtractedlosvds}
\centering
\includegraphics{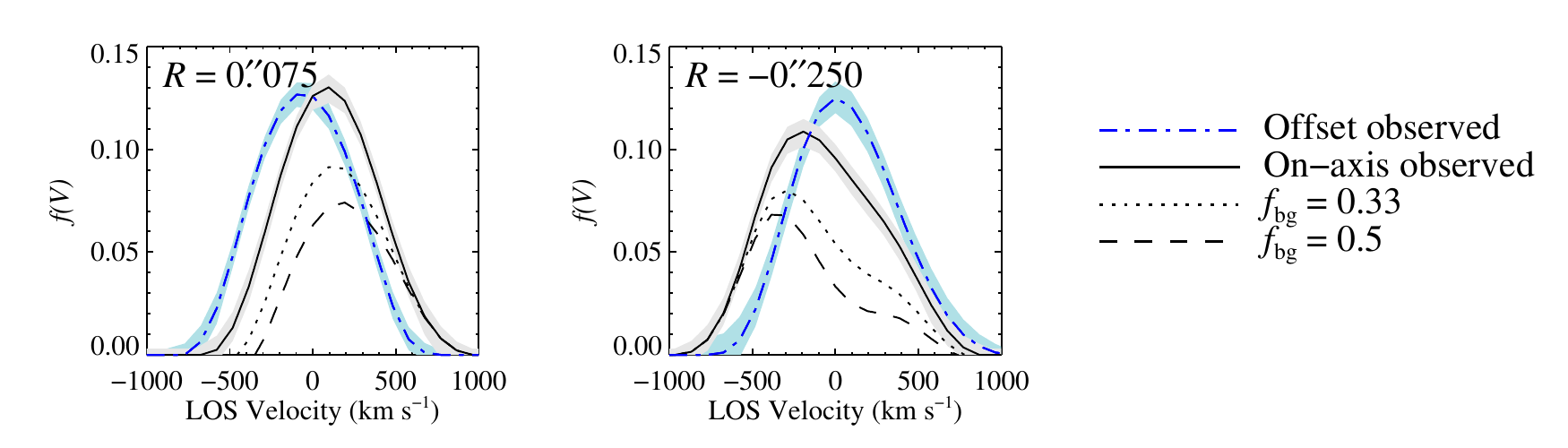}
\caption{\emph{HST}/STIS line-of-sight velocity distributions (LOSVDs)
illustrating how the ring cannot rotate in one direction, counter to
the sense of the rest of the galaxy.  The left panel comes from
cylindrical coordinate $R = 0\farcs075$, and the right panel comes
from cylindrical coordinate $R = -0\farcs250$, where the negative
indicates that it comes from the southeast side of the galaxy.  These
two were chosen because they had high signal-to-noise for both the
on-axis and offset positions.  In both panels, the observed LOSVDs are
plotted as continuous curves with shaded regions indicating 1$\sigma$
uncertainty regions.  The two other curves are the on-axis LOSVD after
subtracting the offset LOSVD, weighted by $f_\mathrm{bg}$.  Thus the
dotted and dashed curves show the LOSVD of the ring only after a
kinematic decomposition, assuming that the rest of the galaxy (other
than the ring) makes up a fraction $f_\mathrm{bg}$ of the total light
in the given bin.  It can be seen that there is still substantial
counter-rotation in just the decomposed ring LOSVD.}
\kgfigstarend{subtractedlosvds}{Decomposed LOSVDs of the stellar ring.}

\subsection{Self-gravity and stability of the ring}

Our analysis of the ring in NGC 3706 raises several questions about 
the stability and longevity of this system.   We consider
(i) the vertical self-gravity of the ring, (ii) the
misalignment of the ring with the bulge, and (iii) the stability of
the counter-rotation.   For ease of
computation, in the remainder of this subsection we make a few
simplifying assumptions about the bulge and ring.  First, we assume
that the axisymmetric spheroidal iso-density contours of the bulge have constant
axis ratio $b/a = 0.68$ and have mass density at a semimajor axis $a$
of
\beq
\rho(a) = \rho_0 (a / a_0)^{-2}
\eeq
where $\rho_0 = 480\ \msun\,\units{pc^{-3}}$ and $a_0 =
50\ \units{pc}$.  This form is an approximation\footnote{The density
  is actually slightly shallower/steeper than $a^{-2}$ inside/outside
  of $200\,\units{pc}$.} for the results from our photometric decomposition using our best-fit mass-to-light ratio, $\Upsilon_V =
6\,\msun\,\lsunv^{-1}$.  We assume the ring has constant surface
density $\Sigma_0 = 6 \times 10^4\,\msun\,\units{pc^{-2}}$ over its
domain from $r = 20$--70\,pc, giving a total ring mass of 
$M_r = 8.5 \times 10^8\,\msun$.  Although the vertical extent of the
ring is unresolved, we assume that the mass density has the functional
form of an isothermal sheet:
\beq
\rho(z) = \frac{\Sigma_0}{4 z_0} \sech^2\left(\frac{z}{2 z_0}\right),
\eeq
where the scale height, $z_0$, must be small enough to remain
unresolved.  Note that the circular velocity at the radius of the ring
is approximately constant at $v_c \approx 400\,\kms$.

\subsubsection{Self-gravity of the ring}

At the outer edge of the stellar ring ($r = 70\ \units{pc}$), the
enclosed mass is made up of the black hole ($M = 6 \times
10^8\,\msun$), the bulge stars ($M_b \approx 8 \times 10^8\,\msun$),
the dark matter halo ($M_\mathrm{DM} = 3 \times 10^4\,\msun$), and the
ring itself ($M_r \approx 8.5 \times 10^8\,\msun$).  The ring's
flatness raises the question of whether or not it is vertically
self-gravitating, i.e., whether the vertical force is dominated by
self gravity instead of gravitational forces from the black hole,
bulge, and dark matter halo.  We examine this question by considering
the ring's \citet{1964ApJ...139.1217T} $Q$ parameter, which typically
is $Q \sim 1$ for a vertically self-gravitating ring or disk embedded
in a spherical stellar system that dominates the radial force.

Toomre's $Q$ for stellar dynamical systems is given by
\beq
Q = \frac{\sigma_r \kappa}{3.36 G \Sigma_0},
\label{toomreq}
\eeq
where $\kappa$ is the epicyclic frequency.  At the outer edge of the
ring, the rotation curve is roughly flat (Fig.\ \ref{f:spec}) so that
$\kappa \approx 1.4 \Omega = 1.4 v_c / r =
2.6\times10^{-13}\,\units{s^{-1}}$.  Thus, $Q \approx 1.3$, and it is
vertically self-gravitating.

\subsubsection{Survival of a ring misaligned with the bulge}

Because the ring is misaligned with respect to the symmetry plane of
the axisymmetric bulge by an angle $i = 42^\circ$, the ring is 
subject to torques, which will cause the ring to precess, and dynamical 
friction, which can damp the misalignment.

The precession timescale of the misaligned ring is $\tau = 
(L \sin{i}) / N_{rb}$, where $L$ is the total angular momentum 
of the ring and $N_{rb}$ is the torque on the ring by the bulge.  
The total angular momentum of the ring about its rotation axis can 
be approximated by taking the entire ring to be at a radius $r_r = 
50\,\units{pc}$ and rotating at a speed $v_\phi = 500\,\kms$ so 
that $L = \frac{1}{3} v_\phi r_r$, where the factor of $1/3$ comes 
from the fact that $\sim 1/3$ of the stars are counter rotating.  
The torque on a ring of mass $M_r$ by the flattened bulge is
\beq
N_{rb} = \frac{5 \pi}{2} G M_r \rho_0 a_0^2 \sin{(2i)} f(c) = M_r n_{br},
\eeq
where we have defined $n_{br}$ for convenience below and 
$f(c)$ is a function of the flattening, written in terms of $c^2 \equiv (a^2/b^2) - 1$:
\beq
f(c) = \frac{3}{2}\frac{c - \arctan{c}}{c^{3}} - \frac{1}{2}\frac{\arctan{c}}{c}.
\eeq
For
$b/a = 0.68$, $f(c) \approx -0.08$.  The precession timescale is
then $\tau \approx 2 \times 10^{6}\,\units{yr}$.

The calculation above assumes that the ring is rigid, which is a 
reasonable approximation if the mutual gravitational torques between 
the ring elements are strong enough to enforce rigid precession. 
  To estimate the accuracy of this approximation, we simplify by splitting the ring into an inner
and outer ring with masses, $M_i = 2 \times 10^{8}\ \msun$ and $M_o =
6.5 \times 10^{8}\ \msun$, and radii $r_i = 20\ \units{pc}$ and $r_o =
70\ \units{pc}$, respectively.  The torque on each ring is then
\beq
N_{i,o} =  M_{i,o} n_{rb}
\eeq
so that the difference in bulge torques is
\beq
N_\Delta = N_o - N_i = (M_o - M_i) n_{rb}.
\label{torquediff}
\eeq
The mutual torque between the two rings when misaligned from each
other by a small angle $\psi$ is
\beq
N_{rr} =  \frac{3}{8} \frac{G M_i M_o}{r_o} \frac{r_i^2}{r_o^2} \sin{2\psi}.
\label{ringselftorque}
\eeq
Combining Eqs. (\ref{torquediff}) and (\ref{ringselftorque}) shows
that the torques balance for $\psi \approx 1^\circ $, which represents 
the amount of warping required to maintain rigid precession; the 
smallness of the warp confirms that the approximation of rigid 
precession is accurate.

The ring--bulge system in NGC 3706 is
analogous to misaligned or warped galactic disks in dark matter halos.
Dynamical friction from the halo damps
warps in galactic disks, causing them to align on very short
timescales, $\tau < 10^{8}\,\units{yr}$ \citep{1995ApJ...442..492D,
  1995MNRAS.275..897N}.  If the same principles apply, the dynamical
friction timescales would be much shorter for the NGC 3706 ring, which
has an orbital period of $\sim 10^6\,\units{yr}$.  Thus the ring
should settle quickly to the symmetry plane of the bulge, and either we are 
 seeing it shortly after creation, or there is another process maintaining the misalignment.

The presence of counter-rotating stars may excite, rather than damp,
the misalignment.  \citet{1995MNRAS.275..897N} found that galactic
warps may be excited by halos but only if the galactic disk and halo
are rotating in opposite senses.  However, given that only about 
one third of the stars counter-rotate, it seems probable that the 
inclination will still damp faster than it would excite.

\subsubsection{Stability of the counter-rotating stream}

The presence of counter-rotating stars raises the possibility that the
ring is subject to a two-stream instability. The physics of the
two-stream instability is more subtle in gravitating systems than in
electrostatic plasmas: in contrast to a homogeneous plasma, a
homogeneous gravitating system is already unstable at wavelengths
larger than the Jeans length, so a two-stream instability can only be
said to be present if the system is less stable than it would have
been if the relative velocity of the two streams were zero (and for
Maxwellian streams the opposite is true; see
\citealt{1967rta2.book..131L} and
\citealt{1987AJ.....94...99A}). \citet{1994ApJ...425..530S}
investigated the stability of self-gravitating disks in which half the
particles were counter-rotating. They found a variety of outcomes
depending on the radial velocity dispersion, ranging from apparent
stability, to buckling instability, to in-plane lopsided ($m=1$)
instabilities. The nature of instabilities in counter-rotating disks
also changes in the Keplerian regime, where the gravitational
potential is dominated by a central point mass: in this case
counter-rotating stars are often unstable to a secular $m=1$
instability that leads to a precessing lopsided disk
\citep{2012MNRAS.423.2083T}. The ring in NGC 3706 is in neither the radially
self-gravitating nor the Keplerian regime since its estimated mass is
similar to the mass of the black hole (see Section \ref{multitidal}).
There is no evidence of asymmetry in the ring, which indicates either
that the ring is axisymmetric, or the $m=1$ component of any
distortion is small, or that there is an $m=1$ component with symmetry
axis along the line of sight. Given the complexity of the dynamics of
counter-rotating rings and the limited accuracy with which we can
determine the velocity distribution of the ring stars, we can make no
definitive statements about stability.

\subsection{Speculation on the ring's formation}
\label{speculation}
The presence of counter-rotating stars in the stellar ring is
unexpected, though not without precedent (Section \ref{comparison}).
Here we speculate on a number of potential ways in which such a system
could have formed.

\subsubsection{Multiple tidal disruptions}
\label{multitidal}
If the creation
of the stellar ring is through dissipative accretion, as argued by
\citet{2002AJ....124.1975L}, it would have to be through at least two
accretion events.  At the inferred mass-to-light ratio of $\Upsilon_V
= 6\ \msun\ \lsunv^{-1}$, the estimated mass of the ring is
approximately $7.8 \times 10^{8}\ \msun$, or approximately equal to
the mass of the black hole.  The stellar mass of the ring could be
supplied by two dwarf galaxies or several tens of stellar clusters.
If the coplanarity of the counter-rotating stars in the stellar ring
is merely a chance alignment of independent disruption events, then it
would be far more likely for it to happen with just two stellar
systems, which would require the two accreted systems to be dwarf galaxies.  
The typical central stellar density
of a dwarf galaxy is $\rho_* \sim 1\ \msun\ \units{pc^{-3}}$.
Assuming that tidal disruption of a stellar system by a black hole of
mass $M$ happens roughly at a distance
\beq
r_d \sim \left(\frac{3 M}{4 \pi \rho_*}\right)^{1/3},
\eeq
dwarf galaxies would be disrupted at distances of $\sim
500\ \units{pc}$, much larger than the size of the ring ($\sim
20$--$75\ \units{pc}$).  Thus, it would require subsequent dynamical
evolution for the tidal debris to shrink to the observed size of the
ring, which seems implausible.  The typical central stellar density of
a globular cluster, however, is $\rho_* \sim
10^{4}\ \msun\ \units{pc^{-3}}$, corresponding to a tidal disruption
distance of $r_d \sim 25\ \units{pc}$, or about the size of the
stellar ring.  So tidal disruption of stellar clusters
self-consistently explains the size of the ring but requires many such
disruptions.  The presence of older globular cluster stars could also
explain the relatively redder color of the central
1\arcsec\ \citep{1994MNRAS.270..523C}.  The chance superposition of
several tens of globular cluster disruptions is so unlikely as to be
precluded.  If, however, there is a preferred plane in the potential
of the galaxy, such as a large-scale disk, it could preferentially
feed stellar clusters towards the central black hole on orbits with
aligned or anti-aligned angular momenta.  We note that this
speculation is not supported by any evidence for a disk, and the misalignment 
of the ring with respect to the symmetry axis of the galaxy on larger 
scales makes this less likely.  

\subsubsection{Multiple gas accretion events}
Another possible explanation is that the ring of stars formed in situ
from gas accreted in two separate mergers.  In such a scenario, the
gas would funnel to the center of the galaxy and subsequently form a
filled-in disk of stars as has been seen in other cases
\citep[e.g.,][]{1989ARA&A..27..235K, 1995AJ....110.2622L,
  2005AJ....129.2636K}.  In order to form a counter-rotating
structure, the first batch of accreted gas would have to form stars
before gas from the second accretion event reached the nucleus.  Note
that this scenario is distinct from the scenario of bar-driven secular formation
of a nuclear disk with rings at the inner Linblad and ultraharmonic
resonances of the tumbling triaxial potential seen in NGC 4750
\citep{1998MNRAS.298..267V}.  There is an established connection
between bars and $\sim\units{kpc}$-scale nuclear rings
\citep[e.g.,][]{1996FCPh...17...95B}. There are cases where
larger-scale nuclear rings are misaligned with respect to the bar
structure \citep[]{1995AJ....110.1588B}, but it is not clear whether
the observed misalignment between stellar ring and galaxy isophotes
seen in NGC 3706 could occur at scales of $\sim10$--$100\ \units{pc}$.
The observations present a number of constraints that can be used to
determine qualitatively whether multiple gas accretion events can make
the observed ring.
The stellar ring has roughly the same mass as the central black hole,
as has been seen in other nuclear disks thought to come from gas
accretion events \citep{1998MNRAS.300..469S}.  So the amount of mass
we infer for the ring is reasonable for it to have come from gas
accretion.
A formation by multiple gaseous accretions would also predict a
different color from the rest of the galaxy, as is observed.
Note that the formation of the ring with multiple gas accretion events
still requires either a preferential plane of the potential to funnel
gas into the same planar orbit or else a coincidence in having the two
accretion events come from the same plane with opposite rotational
senses.  This hypothesis is made more likely by the prevalence of
nuclear disks in other early type galaxies.

\subsubsection{Triaxial to axisymmetric potential}
An interesting, though perhaps unlikely, way to make a
counter-rotating disk is by having the initial disk in a triaxial
potential that becomes axisymmetric.  Box orbits in the triaxial
potential become loop orbits in the axisymmetric potential in either
prograde or retrograde directions \citep{1994ApJ...420L..67E}.  If
there is also a rotating azimuthal potential well, then there will be
an asymmetry in the prograde--retrograde distribution of the resulting
loop orbits as is seen in NGC 3706.

\subsubsection{Resonant capture}
Another mechanism for forming a counter-rotating disk is capture in
the \citet{1981MNRAS.196..455B} resonance, where the rate of
precession of the angular momentum vectors of the disk stars equals
the pattern speed of some non-axisymmetric perturbation, for example
the apsidal precession rate of a binary black hole.  Stars captured in
this resonance can be levitated from prograde to polar and then
retrograde orbits if the pattern speed sweeps from negative to
positive values \citep{2000MNRAS.319....1T}.  The black hole binary
could also have scattered away the inner regions of what was
previously a filled-in disk to turn it into a ring.  In all these 
scenarios, we are resorting to improbable events but this is to be 
expected since the NGC 3706 ring is unusual.

\subsection{Comparison to other counter-rotating systems}
\label{comparison}
NGC 3706 is not the first example of a counter-rotating stellar
system, and here we mention some others.  NGC 4550 is an S0 galaxy
possessing a stellar disk with two spatially coexistent kinematic components  \citep{1992ApJ...394L...9R, 1994AJ....108..456R}.  Not only are the two kinematic components rotating in opposite directions from each other, but they have different stellar
metalicities and ages, precluding a common origin
\citep{2012MNRAS.tmp..103J, 2012arXiv1210.7807C}.  At smaller scales,
the cores of early-type galaxies also show evidence of distinct
rotational components.  Out of a sample of 260 early-type galaxies, as
part of the ATLAS$^\textrm{3D}$ project, \citet{2011MNRAS.414.2923K}
identified 11 as having a kinematically distinct core, 8 as having
counter-rotating cores, and 11 as having two off-center symmetric
peaks in velocity dispersion, separated by at least $0.5 R_e$.  At
scales comparable to the stellar ring in NGC 3706, there is an
eccentric disk in M31 \citep{1993AJ....106.1436L}, which appears to
require a black hole to maintain the eccentricity \citep{tremaine95}.
\citet{2012arXiv1207.1108K} argue from $N$-body simulations that the
eccentric disk was created from counter-rotating stellar clusters,
which is analogous to our hypothesized formation for the stellar ring
in NGC 3706, though the stellar mass in the eccentric M31 disk is much smaller.  At
even smaller scales, within $\sim0.5\ \units{pc}$ of the Galactic center,
there appear to be two young stellar disks that rotate in different
directions \citep{2003ApJ...590L..33L, 2006ApJ...643.1011P,
  2009ApJ...690.1463L, 2009ApJ...697.1741B}.  More recently,
\citet{2013MNRAS.tmp.1094L} report the presence of counter-rotating
orbits in the nuclear star cluster in FCC 277 based on high-spatial
resolution IFU spectra and axisymmetric Schwarzschild modeling.  There
is, however, no direct analogy between the
counter-rotating stellar ring in NGC 3706 and the above examples.

In the end, however unexpected, we have found it necessary to include
a stellar ring with rotation in both directions in order to explain
the observed $|V|/\sigma$ and LOSVD of the ring in NGC 3706.  Based on our mass modeling, the ring
is located at the tidal disruption radius of a stellar cluster for our
black hole mass $M = (6^{+0.7}_{-0.9}) \times 10^8\ \msun$.  While
this coincidence argues for a disruptive encounter between the
progenitor of the stellar ring and the black hole, it requires many
tens of dissipative encounters to make up the entire mass of the ring
and some method for it to be aligned into a plane different from the
galaxy's symmetry plane.  Instead, it is probably more likely that
two, unrelated, merger events led to the funneling of gas to the
center of the galaxy and---after forming stars---resulted in a disk of
stars, some of which are counter-rotating.  A binary black hole could
then scatter the inner stars, turning the disk into a ring, or even be
responsible for the counter-rotation itself.

\hypertarget{ackbkmk}{}%
\acknowledgements 
\bookmark[level=0,dest=ackbkmk]{Acknowledgments}

KG acknowledges support provided by the National
Aeronautics and Space Administration through Chandra Award Number
GO0-11151X issued by the Chandra X-ray Observatory Center, which is
operated by the Smithsonian Astrophysical Observatory for and on
behalf of the National Aeronautics Space Administration under contract
NAS8-03060. This material is based upon work supported by the National
Science Foundation under Grant No.\ AST-1107675.  
ST acknowledges support from NASA grant NNX11AF29G.

This research has made use of the NASA/IPAC Extragalactic Database
(NED) which is operated by the Jet Propulsion Laboratory, California
Institute of Technology, under contract with the National Aeronautics
and Space Administration.  
This research has made use of NASA's Astrophysics Data System.

\bibliographystyle{apjads}
\hypertarget{refbkmk}{}%
\bookmark[level=0,dest=refbkmk]{References}
\bibliography{gultekin}

\label{lastpage}
\end{document}